\documentclass[12pt]{article}
\usepackage{amssymb, amsmath}

\allowdisplaybreaks

\newtheorem{theorem}{Theorem}
\newtheorem{lemma}[theorem]{Lemma}
\newtheorem{proposition}[theorem]{Proposition}

\newtheorem{remark}[theorem]{Remark}
\newtheorem{example}[theorem]{Example}

\newenvironment{proof}{{\em Proof.} }{\medspace}

\newcommand{\qed}{\hfill$\square$}
\newcommand{\PSPACE}{\mathrm{PSPACE}} 
\newcommand{\EXPSPACE}{\mathrm{EXPSPACE}}
\newcommand{\N}{\mathbb{N}}

\newcommand{\B}{\mathbb{B}}

\renewcommand{\phi}{\varphi}

\newcommand{\eval}{\mathrm{eval}}
\newcommand{\pexp}{\mathrm{exp}}
 \newcommand{\leftp}{\mathrm{l}}
 \newcommand{\rightp}{\mathrm{r}}
\newcommand{\head}{\mathrm{head}}
\newcommand{\Head}{\mathrm{Head}}
\newcommand{\body}{\mathrm{body}}
\newcommand{\Body}{\mathrm{Body}}
\newcommand{\tail}{\mathrm{tail}}
\newcommand{\Tail}{\mathrm{Tail}}

\newcommand{\RLangA}{P}

\newcommand{\invol}{\overline{\,^{\,^{\,}}}}
\newcommand{\atab}{\hspace*{10mm}}
\newcommand{\OO}{{\mathcal O}}

\newcommand{\ob}{\bar{b}}
\newcommand{\oa}{\bar{a}}
\newcommand{\oc}{\bar{c}}
\newcommand{\od}{\bar{d\,\!}}
\newcommand{\ox}{\overline{X}}
\newcommand{\oy}{\overline{Y}}

\def\uwv #1{(u_{#1},w_{#1},v_{#1})}
\def\Fl #1#2{\uwv{#1} \cdots \uwv{#2}}

\begin{document}

\title{The existential theory of equations with rational
constraints in free groups is $\PSPACE$--complete}
\author{Volker Diekert$^{1}$,
   Claudio Guti{\'e}rrez$^{2}$,
   Christian Hagenah$^{1}$} 
\date{ \small
	${^{1}}$ Inst.~f{\"u}r Informatik, Universit{\"a}t Stuttgart, \\
    Breitwiesenstr.~20-22, D-70565 Stuttgart  \\
        diekert@informatik.uni-stuttgart.de \;\;
	christian@hagenah.de \\
    ${^{2}}$
     Depto. de Ciencias de la Computaci{\'o}n, 
	 Universidad de Chile, \\
     Blanco Encalada 2120, Santiago, Chile \\
	cgutierr@dcc.uchile.cl
}
\maketitle

\paragraph{ACM Classification} 
F.2. Analysis of Algorithms and Problem Complexity,
F.2.2. Computation on Discrete Structures,
F.4. Mathematical Logic and Formal Languages. \\
{\bf Subject Descriptor} Equations in Free Groups.

\begin{abstract}
It is known that the existential theory of equations in free groups is
decidable. This is a famous result of Makanin. 
On the other hand it has been shown that the scheme of his
 algorithm is not primitive recursive.  
In this paper we present an algorithm that works in polynomial space,
even in the more general setting where each variable has a rational
constraint, that is, the solution has to respect a specification given
by a regular word language.  
Our main result states that the existential
theory of equations in free groups with rational constraints is
$\PSPACE$--complete.  We obtain this result as a corollary of the
corresponding statement about free monoids with involution.
\end{abstract}

\section{Introduction}\label{sec:intro}

 Around the 1980's a great progress was achieved on the algorithmic
decidability of elementary theories of free monoids and groups.  In
1977 Makanin \cite{mak77} proved that the existential theory of
equations in free monoids is decidable by presenting an algorithm
which solves the satisfiability problem for a single word equation
with constants. In 1983 he extended his result to the more complicated
framework in free groups \cite{mak83a}. In fact,
using a result by Merzlyakov \cite{merz66} he also showed 
that the positive theory of equations in free groups is decidable
\cite{mak84}, and 
Razborov was able to
give a description of the whole solution set \cite{raz84}. The
algorithms of Makanin are very complex: For word equations the running
time was first estimated by several towers of exponentials and it took
more than 20 years to lower it down to the best known bound for
Makanin's original algorithm, which is to date $\EXPSPACE$
\cite{gut98focs}.  For solving equations in free groups Ko{\'s}cielski
and Pacholski \cite{kp98} have shown that the scheme proposed by
Makanin is not primitive recursive.

In 1999 Plandowski invented another method for solving word
equations and he showed that the satisfiability problem for word
equations is in $\PSPACE$, \cite{pla99focs}.  One ingredient of his
work is to use data compression to reduce the exponential space to
polynomial space.  The importance of data compression was first
recognized by Rytter and Plandowski when applying {L}empel-{Z}iv
encodings to the minimal solution of a word equation \cite{pr98icalp}.
Another important notion is the definition of an $\ell$-factorization
of the solution being explained below.  Guti{\'e}rrez extended
Plandowski's method to the case of free groups, \cite{gut2000stoc}.
Thus, a non-primitive recursive scheme for solving equations in free
groups has been replaced by a polynomial space bounded algorithm.
Hagenah and Diekert worked independently in the same direction and
using some ideas of  Guti{\'e}rrez they obtained a result which 
includes the presence of rational constraints. This appeared as extended
abstract in \cite{dgh2001} and also
as a part 
 of the PhD-thesis of Hagenah \cite{hagenahdiss2000}.
 
 The present paper is a journal version of
 \cite{dgh2001,gut2000stoc}. It shows that the existential theory of
 equations in free groups with rational constraints is
 $\PSPACE$--complete.  Rational constraints mean that a possible
 solution has to respect a specification which is given by a regular
 word language.  The idea to consider regular constraints for word
 equations goes back to Schulz \cite{sch91} who also pointed out the
 importance of this concept, see also \cite{dmm99tcs,gv97icalp}.
 The
 $\PSPACE$--completeness for the case of word equations with regular
 constraints has been stated by Rytter already, as cited in
 \cite[Thm.~1]{pla99focs}. 

 Our proof reduces the case of equations with rational constraints in free groups
to the case equations with regular constraints in free monoids with involution,
which turns out to be the central object.
(Makanin uses the notion of ``paired alphabet'', but a main difference is that 
he considered ``non contractible'' solutions only, whereas we deal
with general solutions and, in addition, we have constraints.)
During our work we extend the method of
\cite{pla99focs} such that it copes with the involution and the method
of \cite{gut2000stoc} such that it copes with rational constraints.
The first step is a reduction to the satisfiability problem of a
single equation with regular constraints in a free monoid with
involution. In order to avoid an exponential blow-up, we do not use a
reduction as in \cite{mak84}, but a simpler one. In particular, we can
handle negations simply by  positive rational constraints. In the
second step we show that the satisfiability problem of a single
equation with regular constraints in a free monoid with involution is
still in $\PSPACE$. This part is rather technical and we introduce
several new notions like base-change, projection, partial solution, and free
interval. The careful handling of free intervals is necessary because
of the constraints. In some sense this is the only additional
difficulty which we will meet when dealing with constraints. After
these preparations we can follow Plandowski's method.
Throughout we shall use many of the deep ideas which were presented 
in \cite{pla99focs}, and apply them in a different setting.
 Hence, as we cannot use
Plandowski's result as a black box, we have to go through the whole
construction again. As a result our paper is (involuntarily)
self-contained, up to standard knowledge in combinatorics on words and
linear Diophantine equations.

\section{Free Groups and their Rational Subsets}

Let $\Sigma$ be a finite alphabet. By $F(\Sigma)$ we denote the free group
over $\Sigma$. Elements of $F(\Sigma)$ can be represented by words in
$(\Sigma \cup \overline{\Sigma})^{*}$, where
$\overline{\Sigma}= \{\, \overline{a} \mid a \in \Sigma \,\}$. We read
$\overline{a}$ as $a^{-1}$ in $F(\Sigma)$ and we use the convention that
$\overline{\overline{a}}=a$.
Hence the set $\Gamma=\Sigma \cup \overline{\Sigma}$ is equipped with an
involution
 $\invol:\Gamma \to \Gamma$;
the involution is extended to $\Gamma^{*}$ by
$\overline{a_1 \cdots a_n} = \overline{a_n} \cdots \overline{a_1}$ for
$n \geq 0$ and $a_i \in \Gamma$, $1 \leq i \leq n$.
The empty word as well as the unit element in other monoids
is denoted by $1$.  By $\psi : \Gamma^* \to
F(\Sigma)$ we denote the canonical homomorphism.
A word $w \in \Gamma^*$ is {\em freely reduced\/}, if it
contains no factor of the form $a \overline{a}$ with $a \in \Gamma$.
The
reduction of a word $w \in \Gamma^*$ can be computed by using the Noetherian
and confluent rewriting system $\{\,a\overline{a}\rightarrow 1 \mid a \in \Gamma\,\}$.
For $w \in \Gamma^*$ we denote by $\widehat w$ the freely reduced word which denotes
the same group element in $F(\Sigma)$ as $w$.
Hence, $\psi(u)=\psi(v)$ if and only if $\widehat u=\widehat v$ in $\Gamma^*$.

The class of {\em rational languages} in $F(\Sigma)$ is inductively defined
as follows: Every finite subset of $F(\Sigma)$ is
rational.
If
$\RLangA_1,\RLangA_2 \subseteq F(\Sigma)$ are rational, then
$\RLangA_1 \cup \RLangA_2$, $\RLangA_1 \cdot \RLangA_2$, and
$\RLangA_1^*$ are rational.  Hence, $\RLangA \subseteq F(\Sigma)$ is
rational if and only if $\RLangA=\psi(\RLangA')$ for some regular
language $\RLangA' \subseteq \Gamma^*$.\footnote{We 
follow the usual convention to call a rational 
subset of a free monoid {\em regular}.
This convention is due to Kleene's Theorem
stating that regular, rational, and recognizable
have the same meaning in free monoids. But in free
groups these notions are different and we have 
to be more precise.} In particular, we can use
a non-deterministic
finite automata over $\Gamma$ for specifying   rational group
languages over $F(\Sigma)$.

The following proposition  is due to M.~Benois \cite{ben69}, 
see also \cite[Sect. III. 2] {berstel79}.

\begin{proposition} \label{benois}
Let $\RLangA' \subseteq \Gamma^*$ be a regular language and
$\RLangA=\psi(\RLangA') \subseteq F(\Sigma)$. Then we effectively find a
regular language $\widetilde{\RLangA}'\subseteq \Gamma^*$ such that
$\widetilde{\RLangA}'=\{\,\widehat w \in \Gamma^* \mid \psi(w) \not\in \RLangA
\,\}$.
Hence,
the complement of $\RLangA$ is the rational group language
$\psi(\widetilde{\RLangA}')$
and the family of rational group languages is an effective Boolean algebra.
\end{proposition}

\begin{proof}
  (Sketch) Using the same state set (and some additional transitions
which are labeled with the empty word) we can construct (in polynomial
time) a finite automaton which accepts the following language
$$\RLangA''=\{\,v \in \Gamma^* \mid \exists u\in \RLangA':
  u \stackrel{*}{\rightarrow} v  \,\}$$
where $u \stackrel{*}{\rightarrow} v$ means that $v$ is a descendant of $u$ by
the convergent rewriting system $\{\,a \overline{a} \rightarrow 1 \mid a \in
\Gamma\,\}$.
Then we complement $\RLangA''$ with respect to
$\Gamma^*$; and we build the intersection
with the regular set of freely reduced words.\qed
\end{proof}

\section{The Existential Theory}

In the following $\Omega$ denotes a finite set of variables (or unknowns) and we
let $\invol:\Omega \to \Omega$ be an involution without
fixed points. Clearly, if $X \in \Omega$ has an interpretation in $F(\Sigma)$,
then we read $\overline{X}$ as  $X^{-1} \in F(\Sigma)$.

The {\em existential theory of equations with rational constraints in free groups} is
inductively defined as follows. Atomic formulae are either of the form $W=1$,
where $W \in (\Gamma \cup \Omega)^*$ or of the form $X \in \RLangA$,
where $X$ is in 
$\Omega$ and $\RLangA \subseteq F(\Sigma)$ is a rational language. A propositional
formula is build up by atomic formulae using negations, conjunctions and
disjunctions. The existential theory refers to closed existentially quantified
propositional formulae which evaluate to {\em true\/} over $F(\Sigma)$.

\begin{theorem} \label{theo1}
The following problem is $\PSPACE$--complete.

INPUT: A closed existentially quantified propositional formula with rational
constraints in the free group $F(\Sigma)$ for some finite alphabet $\Sigma$.

QUESTION: Does the formula evaluate to {\em true\/} over $F(\Sigma)$?
\end{theorem}

The $\PSPACE$--hardness follows {}from a result of Kozen
\cite{koz77}, since (due to the constraints) the {\em empty
intersection} problem of regular sets can easily be encoded in the
problem above. The same argument applies to Theorems~\ref{theo2} and
\ref{theo3} below and therefore the $\PSPACE$--hardness is not
discussed further in the sequel: We have to show the inclusion in $\PSPACE$,
only.

The $\PSPACE$ algorithm for solving Theorem~\ref{theo1} will be
described by a (highly) non-deterministic procedure. We will make sure
that if the input evaluates to true, then at least one possible output
is true. If it evaluates to false, then no (positive) output is
possible. By standard methods (Savitch's Theorem) such a procedure can
be transformed into a polynomial space bounded deterministic decision
procedure, see any textbook on complexity theory,
e.g.~\cite{hu79,pap94}.

We start the procedure as follows. Using the rules of DeMorgan we may assume
that there are no negations at all, but the atomic formulae are now of the
either form: $W=1$, $W \neq 1$, $X \in \RLangA$, $X \not\in \RLangA$ with
$W\in (\Gamma \cup \Omega)^*$, $X \in \Omega$, and $\RLangA \subseteq F(\Sigma)$ 
rational.\footnote{The reason that we keep $X \not\in \RLangA$ instead of
$X \in \widetilde \RLangA$ where $\widetilde \RLangA = F(\Sigma) \setminus \RLangA$ is
that the complementation may involve an exponential blow-up of the state space;
this has to be avoided.}

The next step is to replace every formula $W \neq 1$ by
$$\exists X: WX=1 \wedge X \not\in \{1\},$$
where X is a fresh variable, hence we can put $\exists X$ to the front. Now we
eliminate all disjunctions. More precisely, every subformula of type $A \vee B$
is non-deterministically replaced either by $A$ or by $B$. At this stage the
propositional formula has become a conjunction of formulae of type $W=1$,
$X \in \RLangA$, $X \not\in \RLangA$ with $W \in (\Gamma \cup \Omega)^*$,
$X \in \Omega$, and $\RLangA \subseteq F(\Sigma)$ rational.

We may assume that $|W|\geq 3$, since
if $1 \leq |W| < 3$, then we may replace $W=1$ by $Wa\overline{a}=1$
for some $a \in \Gamma$. For the following it is convenient to assume
that $|W|=3$ for all subformulae $W=1$. This is also easy to achieve. As
long as there is a subformula $x_1 \cdots x_k = 1$, $x_i \in \Gamma
\cup \Omega$ for $1 \leq i \leq k$ and $k \geq 4$, we replace it by the
conjunction
$$\exists Y: x_1 x_2 Y = 1 \wedge \overline{Y} x_3 \cdots x_k=1,$$
where $Y$ is a fresh variable and $\exists Y$ is put to the front, and
then proceed recursively.
This finishes the first phase. The output of this phase is a system of atomic
formulae of type $W=1$, $X \in \RLangA$, $X \not\in \RLangA$ with
$W \in (\Gamma \cup \Omega)^3$, $X \in \Omega$, and $\RLangA \subseteq
F(\Sigma)$ rational.

At this point we switch to the existential theory of equations with
regular constraints in free monoids where these monoids 
have an  involution.  Recall that $X
\in \RLangA$ (resp. $X \not\in \RLangA$) means in fact $X \in
\psi(\RLangA')$ (resp. $X \not\in \psi(\RLangA')$) where $\RLangA'
\subseteq \Gamma^*$ is a regular word language specified by some
finite non-deterministic automaton. 
Using $\psi$-symbols we obtain an interpretation over
$(\Gamma^*,\invol)$ without changing the truth value by
replacing syntactically each subformula $X \in \RLangA$ (resp. $X
\not\in \RLangA$) by $\psi(X) \in \psi(\RLangA')$ (resp. $\psi(X)
\not\in \psi(\RLangA')$) and by replacing each subformula $W=1$ by
$\psi(W)=1$.

We keep the interpretation over words,
but we eliminate now all occurrences of $\psi$ again. We begin with the
occurrences of $\psi$ in the constraints. Let $\RLangA' \subseteq
\Gamma^*$ be regular being accepted by some finite automaton with
state set $Q$.  As stated in the in the first part of the proof of
Proposition~\ref{benois}, we construct a finite automaton, using the
same state set, which accepts the following language
$$\RLangA''=\{\,v \in \Gamma^* \mid \exists u\in \RLangA':
  u \stackrel{*}{\rightarrow} v  \,\}.$$
In particular, $\psi(\RLangA') = \psi(\RLangA'')$ and
$\widehat \RLangA \subseteq \RLangA''$ where
$\widehat \RLangA = \{\,\widehat u \in \Gamma^* \mid u \in \RLangA'\,\}$.

We replace all positive atomic subformulae of the form
$\psi(X) \in \psi(\RLangA')$ by $X \in \RLangA''$. A simple reflection shows that
the truth value has not changed since we can think of $X$ of being a freely reduced
word. For a negative formulae $\psi(X) \not\in \psi(\RLangA')$ we have to be a
little more careful. Let $N \subseteq \Gamma^*$ be the regular set of all
freely reduced words. The language $N$ is accepted by a deterministic finite automaton
with $|\Gamma|+1$ states. We replace $\psi(X) \not\in \psi(\RLangA')$ by
\[
   X \not\in \RLangA'' \wedge X \in N,
\]
where $\RLangA''$ is as above. Again the truth value did not change.

We now have to deal with the formulae $\psi(xyz)=1$ where $x,y,z \in \Gamma \cup
\Omega$. Observe that the underlying propositional formula is satisfiable over
$\Gamma^*$ if and only if it is satisfiable in freely reduced words. The following
lemma is well-known. Its easy proof is left to the reader.

\begin{lemma} \label{lem1}
Let $u,v,w \in \Gamma^*$ be freely reduced words.  Then we have
$\psi(uvw)=1$ (i.e. $uvw=1$ in $F(\Sigma)$) if and only if there are
words $P, Q, R \in \Gamma^*$ such that $u=PQ$, $v=\overline{Q}R$, and
$w=\overline{R}\,\overline{P}$ holds in $\Gamma^*$.
\end{lemma}

Based on this lemma we replace each atomic subformulae $\psi(xyz)=1$ with $x,y,z
\in \Gamma \cup \Omega$ by a conjunction
$$\exists P \exists Q \exists R: x=PQ \wedge y=\overline{Q}R \wedge
z=\overline{R}\,\overline{P},$$
where $P$, $Q$, $R$ are fresh variables
and the existential block is put to the front. The new existential
formula has no occurrence of $\psi$ anymore. The atomic subformulae
are of the form $x=yz$, $X \in \RLangA$, $X \not\in \RLangA$, where
$x,y,z\in \Gamma \cup \Omega$ and $\RLangA \subseteq \Gamma^*$ is
regular. The size of the formula is linear in the size of the original
formula. Therefore Theorem~\ref{theo1} is a consequence of Theorem~\ref{theo2}.

\section{Free Monoids with Involution}

As above, let $\Gamma$ be an alphabet of constants and $\Omega$ be an alphabet of
variables. There are involutions $\invol:\Gamma \to
\Gamma$ and $\invol:\Omega \to \Omega$. The involution
on $\Omega$ is without fixed points, but  we explicitly allow fixed points for
the involution on $\Gamma$.  
\footnote {Fixed points for
the involution on constants are needed in the proof later anyhow
and this more general setting leads to further applications, \cite{dl2001}}
The involution is extended to
$(\Gamma \cup \Omega)^*$ by
$\overline{x_1 \cdots x_n} = \overline{x_n} \cdots \overline{x_1}$ for
$n \geq 0$ and $x_i \in \Gamma \cup \Omega$, $1 \leq i \leq n$.

{}From now on, all monoids $M$ under consideration are equipped with
an involution $\invol:M \to M$, i.e. we have $\overline{1} = 1$ for
the unit element, $\overline{\overline{x}}=x$, and
$\overline{xy}=\overline{y}\,\overline{x}$ for all $x,y\in M$. A
homomorphism between monoids $M$ and $M'$ is therefore a mapping $h: M
\to M'$ such that $h(1)=1$, $h(xy)=h(x)h(y)$, and
$h(\overline{x})=\overline{h(x)}$ for all $x,y \in M$. The pair
$(\Gamma^*, \invol)$ is called a \emph{free monoid with
  involution}.
\footnote{Note that   $(\Gamma^*, \invol)$ 
  is a free monoid which has an involution, but
  it is not a free object in the category of
  monoids with involution,  as soon as the involution has fixed points.}

The existential theory of equations with regular constraints in free
monoids with involution is based on atomic formulae of type $U=V$
where $U,V \in (\Gamma \cup \Omega)^*$ and of type $X \in \RLangA$
where $X \in \Omega$ and $\RLangA \subseteq \Gamma^*$ is a regular
language specified by some non-deterministic finite automaton.  Again,
a propositional formula is build up by atomic formulae using
negations, conjunctions and disjunctions. The existential theory
refers to closed existentially quantified propositional formulae which
evaluate to {\em true\/} over $(\Gamma^*,\invol)$.

The following statement is the main result of the paper.

\begin{theorem} \label{theo2}
The following problem is $\PSPACE$--complete.

INPUT: A closed existentially quantified propositional formula with regular
constraints in a free monoid with involution over  $(\Gamma, \invol)$.

QUESTION: Does the formula evaluate to {\em true\/} over $(\Gamma^*,\invol)$?
\end{theorem}

The proof of Theorem~\ref{theo2} is in a first step 
(next section) a reduction
to Theorem~\ref{theo3}. The proof of Theorem~\ref{theo3}
will be the essential technical contribution.

\section{From Regular Constraints to Boolean Matrices 
	and a Single Equation}

The first part of the proof is very similar to what we have done above.
By DeMorgan we have no negations and all subformulae are of type
$U=V$, $U \neq V$, $X \in \RLangA$, $X \not\in \RLangA$, where
$U,V \in (\Gamma \cup \Omega)^*$, $X \in \Omega$, and $\RLangA \subseteq \Gamma^*$
is regular.

Since we work over a free monoid $\Gamma^*$ it is easy to handle inequalities
$U \neq V$ where $U,V \in (\Gamma \cup \Omega)^*$. 
We recall it under the assumption $|\Gamma| \geq 2$:
A subformulae $U\neq V$ is replaced by
\[
  \exists X \exists Y \exists Z: \bigvee_{a \neq b}
(U=VaX \vee V=UaX \vee (U=XaY \wedge V=XbZ)).
\]
Making guesses we can eliminate all disjunctions and we obtain a
propositional formula which is a single conjunction over subformulae of type
$U=V$, $X \in \RLangA$, and $X \not\in \RLangA$ where
$U,V \in (\Gamma \cup \Omega)^*$, $X \in \Omega$, and
$\RLangA \subseteq \Gamma^*$ is regular.

By another standard procedure we can replace a conjunction of word
equations over $(\Gamma \cup \Omega)^*$ by a single word equation
$L=R$ with $L,R \in (\Gamma \cup \Omega)^+$. For example,
we may choose a new letter $a$ and then 
we can replace a system 
$L_1 = R_1$, $L_2 = R_2, \ldots , L_k = R_k$ by
$L_1a L_2a \cdots a L_k = R_1a R_2a \cdots a R_k$ and a list
$X \in \Gamma^*$ for all $X \in \Omega$;
this works since $a \not\in \Gamma$. 

 Therefore we may assume 
that our input is given by a single equation
$L=R$ with $L,R \in (\Gamma \cup \Omega)^+$ and by two lists $(X_j \in
\RLangA_j, 1\leq j \leq m)$ and $(X_j \not\in \RLangA_j, m < j \leq
k)$ where $X_j \in \Omega$  and each regular
language $\RLangA_j \subseteq \Gamma^*$ is specified by some
non-deterministic automaton ${\mathcal A}_j = (Q_j, \Gamma, \delta_j,
I_j, F_j)$ where $Q_j$ is the set of states, $\delta_j \subseteq Q_j
\times \Gamma \times Q_j$ is the transition relation, $I_j \subseteq
Q_j$ is the subset of initial states, and $F_j \subseteq Q_j$ is the
subset of final states, $1 \leq j \leq k$.  Of course, a variable $X$
may occur several times in the list with different constraints,
therefore we might have $k$ greater than $|\Omega|$.
The question is whether there is a solution.

  A \emph{solution} is a mapping $\sigma: \Omega \to
\Gamma^*$ 
being extended
to a homomorphism $\sigma: (\Gamma \cup \Omega)^* \to \Gamma^*$ by
leaving the letters from $\Gamma$ invariant  such that the
following conditions are satisfied:
\[
\begin{array}{rcll}
 \sigma(L) & = & \sigma(R), \\
\sigma(\overline{X})& = &\overline{\sigma(X)}&\textrm{for } X \in \Omega,\\
 \sigma(X_j) & \in & \RLangA_j & \textrm{for } 1\leq j \leq m,\\
 \sigma(X_j) & \not\in & \RLangA_j &\textrm{for } m < j \leq k.
\end{array}
\]

 For the next steps it turns out to be more convenient
to work within the framework of Boolean matrices instead of finite automata:
Let $Q$ be the disjoint union of the state spaces $Q_j$, $1 \leq j \leq k$.
We may assume that $Q=\{1,\ldots,n\}$. Let
$\delta = \bigcup_{1 \leq j \leq k} \delta_j$, then
$\delta \subseteq Q \times \Gamma \times Q$ and with each $a \in \Gamma$ we
can associate a Boolean $n \times n$ matrix $g(a) \in \B^{n \times n}$
such that $g(a)_{i,j} = $ ``$(i, a, j) \in \delta$'' for $1 \leq i,j \leq n$.
Since our monoids should have an involution, we
shall in fact work with $2n \times 2n$ matrices. 
Henceforth $M \subseteq \B^{2n \times 2n}$ denotes the following monoid with
involution:
$$M= \{\, \begin{pmatrix}A & 0 \\ 0 & B \end{pmatrix}
	 \mid A,B \in \B^{n \times n} \,\},$$
where $$ \overline{ \begin{pmatrix} A & 0 \\ 0 & B \end{pmatrix}} =
\begin{pmatrix} B^T & 0 \\ 0 & A^T \end{pmatrix}$$
and the operator ${\ }^T$ denotes the transposition. We define a homomorphism
$h: \Gamma^* \to M$ by
$$h(a)=\begin{pmatrix} g(a) & 0 \\ 0 & g(\overline{a})^T \end{pmatrix}
\,\textrm{for}\, a \in \Gamma,$$
where the mapping 
$g: \Gamma \to \B^{n \times n}$ is defined as above. The homomorphism $h$
can be computed in polynomial time and it respects the involution. Now, for
each regular language $\RLangA_j$, $1 \leq j \leq k$ we compute vectors
$I_j, F_j \in \B^{2n}$ such that for all $w \in \Gamma^*$ and $1 \leq j \leq k$
we have the equivalence:
$$w \in \RLangA_j \Leftrightarrow I_{j}^T h(w) F_j =1.$$
Having done these computations we make a non-deterministic guess
$\rho(X) \in M$ for each variable $X \in \Omega$. We verify
$\rho(\overline{X})=\overline{\rho(X)}$ for all
$X \in \Omega$ and whenever there is a constraint of type $X \in \RLangA_j$
for some $1 \leq j \leq m$ (or $X \not\in \RLangA_j$ for some $m<j \leq k$),
then we verify $I_{j}^T \rho(X) F_j=1$, if $1 \leq j \leq m$ (or
$I_{j}^T \rho(X) F_j=0$, if $m<j \leq k$). 

 After these preliminaries, we introduce the formal definition of an
{\em equation $E$ with constraints}: 
Let $d,n\in \N$ and let $M \subseteq
\B^{2n \times 2n}$ be the monoid with involution defined above.  We
consider an equation of  length $d$ over some $\Gamma$ and $\Omega$
with constraints in $M$ being specified by a list $E$ containing the
following items:
\begin{itemize}
\item The alphabet $(\Gamma,\invol)$ with involution.
\item The homomorphism $h: \Gamma^* \to M$ which is specified by a
      mapping $h: \Gamma \to M$ such that
      $h(\overline{a})=\overline{h(a)}$ for all
      $a \in \Gamma$.
\item The alphabet $(\Omega,\invol)$ with involution without fixed points.
\item A mapping $\rho: \Omega \to M$ such that
      $\rho(\overline{X})=\overline{\rho(X)}$ for all
      $X \in \Omega$.
\item The equation $L=R$ where $L,R \in (\Gamma \cup \Omega)^+$ and $|LR|=d$.
\end{itemize}
We will denote this list simply by
\[
	E = (\Gamma, h, \Omega, \rho ; L=R).
\]

A convenient definition for the input size is 
given by $n + d + \log_2(|\Gamma|+|\Omega|)$.
This definition takes into account that there might be constants
or variables with constraints which are not present in the equation.
Recall that $n$ refers to the dimension of the boolean matrices, and
this parameter is part of the input.

A {\em solution} of $E$ is 
a mapping $\sigma: \Omega \to \Gamma^*$ (being extended
to a homomorphism $\sigma: (\Gamma \cup \Omega)^* \to \Gamma^*$
by leaving the letters from $\Gamma$ invariant) such that the following
three conditions are satisfied:
\begin{eqnarray*}
 \sigma(L) & = & \sigma(R), \\
 \sigma(\overline{X}) & = & \overline{\sigma(X)} \,
 \textrm{for all}\, X\in \Omega, \\
 h\sigma(X) & = & \rho(X) \,\textrm{for all}\, X\in \Omega.
\end{eqnarray*}
By the
reduction above, Theorem~\ref{theo2} is a consequence of the next statement
which says that the satisfiability problem of equations with constraints
can be solved in polynomial space.

\begin{theorem} \label{theo3}
The following problem is $\PSPACE$--complete.

INPUT: An equation $E_0$ with constraints
$E_0 = (\Gamma_0, h_0 , \Omega_0, \rho_0 ; L_0=R_0).$

QUESTION: Is there a solution $\sigma : \Omega_0 \to \Gamma_0^*$?
\end{theorem}

For the proof we need an explicit space
bound. Therefore we fix some polynomial $p$ and and
we allow working space $p(n + d + \log_2(|\Gamma|+|\Omega|)$.
An appropriate choice of the polynomial $p$ 
can be calculated {}from the presentation below.
What is important is that the notions of
{\em admissibility\/} being used in the next sections always refer to some fixed polynomials.
The following lemma states that some  basic operations, which
we have to perform several times
can be done in $\PSPACE$.

\begin{lemma} \label{basicops}
The following two problems can be
solved in polynomial space with respect to the
input size $ n + \log ( |\Gamma|)$.

\smallbreak
INPUT: A matrix $A \in M$ and a mapping
$h: \Gamma \to M$.

QUESTION: Is there some $w \in \Gamma^*$ such that $h(w)=A$?
\smallbreak
INPUT: A matrix $A \in M$ and a mapping
$h: \Gamma \to M$.

QUESTION: Is there some $w \in \Gamma^*$ such that $h(w)=A$ and
$w=\overline{w}$?
\end{lemma}
\begin{proof}
  The first question can be solved by guessing a word $w$
letter by letter and calculating $h(w)$. The second question can be
solved since $w=\overline{w}$ implies $w=ua\overline{u}$ for some
$u \in \Gamma^*$ and $a \in \Gamma \cup \{1\}$ with $a=\overline{a}$. Hence 
we can guess $u$ and $a$. During the guess we compute
$B=h(u)$ and then we verify $A=Bh(a)\overline{B}$. \qed
\end{proof}

Here is a first application of Lemma~\ref{basicops}: Assume that an  equation with constraints
$E = (\Gamma, h , \Omega, \rho ; L=R)$
contains in the specification some variable $X$ which does not occur
in $LR\overline{LR}$, then the equation might be unsolvable, simply
because $\rho(X) \not\in h(\Gamma^*)$. However, by
the lemma above
we can test this in $\PSPACE$. If $\rho(X) \in h(\Gamma^*)$,
then we can safely cancel $X$ and $\overline{X}$.
Thus, we put this test in the preprocessing,
and in the following we shall assume that all variables occur
somewhere in
$LR\overline{LR}$. In particular, we may assume $|\Omega| \leq 2|LR|$.

\section{The Exponent of Periodicity}
\label{teop }
 A key step in proving Theorem~\ref{theo3}  
is to find a bound on the exponent of periodicity
in a minimal solution. 
This idea is used in all known algorithms for solving 
word equations in general,
c.f., \cite{mak77,pla99focs}.

Let $w \in \Gamma^*$ be a word. The exponent of periodicity $\pexp(w)$ is
defined by
$$\pexp(w) = \sup \{\,\alpha \in \N \mid \exists u,v,p \in \Gamma^*,p \neq 1:
w=u p^\alpha v\,\}.$$
We have $\pexp(w)>0$ if and only if $w$ is not the empty word. Let
$E = (\Gamma, h, \Omega, \rho, L=R)$
be an equation with constraints. The exponent
of periodicity of $E$ is also denoted by $\pexp(E)$. It is defined by
$$\pexp(E) = \inf\{ \{\,\pexp(\sigma(L))  \mid
\sigma {\rm\,is\,a\,solution\,of\,} E\,\} \cup \{\infty\} \, \}.$$
By definitions we have $\pexp(E) < \infty$ if and only if $E$ is solvable. 
Here we show that the well-known result from word equations
\cite{kp96} transfers to the situation here. The exponent of periodicity of
a solvable equation can be bounded by a singly exponential function. 
Thus, in the following sections we shall assume that if $E_0$
is solvable, then $\pexp(E_0) \in 2^{\OO(d+n\log n)}$.
This is the content of the next proposition.

\begin{proposition}\label{singleexp}
  Let $ E = (\Gamma,h, \Omega, \rho; L=R)$ be an equation with constraints and
  let $\sigma:\Omega \rightarrow \Gamma^*$ be a solution.  Then we
  find effectively a solution $\sigma':\Omega \rightarrow \Gamma^*$
  such that $\exp(\sigma'(L)) \in 2^{\OO(d + n\log n)}$.
\end{proposition}

 The rest of this section is devoted to prove
Proposition~\ref{singleexp}.
Since it follows standard lines, the proof
can be skipped in a first reading.

\begin{proof}
Let $p\in A^{+}$ be a primitive word.  In our setting the definition of 
the {\em $p$-stable normal
  form} of a  word $w \in A^*$ 
depends on the property whether or not $\overline{p}$
is a factor of $p^2$. So we distinguish two cases
and in the following we also write 
$p^{-1}$
for denoting $\overline{p}$. Then, for example, $p^{-3}$
means the same as $\overline{p}^{3}$.

First case: We assume that $\overline{p}$ is not a factor of $p^2$.
The idea is to replace each maximal factor of the form $p^{\alpha}$
with $\alpha \geq 2$ by a sequence $p,\alpha - 2,p$ and each maximal
factor of the form $\overline{p}^{\alpha}$ with $\alpha \geq 2$ by a
sequence $\overline{p},-(\alpha - 2),\overline{p}$.
This leads to the following notion:

The {\em $p$-stable normal
  form} (first kind) of  $w \in A^*$
is a
shortest sequence ($k$ is minimal)
\[
( u_{0},\varepsilon_{1}\alpha _{1},u_{1},\ldots ,
\varepsilon_{k}\alpha _{k},u_{k}) 
\]
such that $k\geq 0$, $u_{0},u_{i}\in A^{*}$, $\varepsilon_{i} \in
\{+1,-1\}$,
$\alpha _{i}\geq 0$
 for
$1\leq i\leq k$, and the following conditions are satisfied:
\begin{itemize}
\item  $w=u_{0}p^{\varepsilon_{1}\alpha _{1}}u_{1}\cdots 
p^{\varepsilon_{k}\alpha _{k}}u_{k}$.

\item  $k=0$ if and only if neither $p^{2}$ 
nor  $\overline{p}^2$ is a  factor of $w.$

\item  If $k\geq 1$, then: 
\begin{eqnarray*} 
u_{0} &\in  &A^{*}p^{\varepsilon_{1}}\setminus A^{*}p^{\pm 2}A^{*}, \\
u_{i} &\in  &( A^{*}p^{\varepsilon_{i+1}}\cap p^{\varepsilon_{i}} A^{*}) 
\setminus A^{*}p^{\pm 2}A^{*}\,
\mbox{ for }1\leq i<k, \\
u_{k} &\in & p^{\varepsilon_{k}}A^{*}\setminus A^{*}p^{\pm 2}A^{*}.
\end{eqnarray*}
\end{itemize}

The $p$-stable normal form of $\overline{w}$
becomes
$$( \overline{u_{k}},-\varepsilon_{k}\alpha _{k},u_{k-1},\ldots 
,-\varepsilon_{1}\alpha _{1},\overline{u_{0}}) .$$

\begin{example} Let $ p = a\overline{a}ba\overline{a} $ with $b \not=\overline{b}$ 
and $w = p^4\overline{b}a\overline{a} p^{-1}a\overline{a}\overline{b}p^{-2}$.
Then the $p$-stable normal form of $w$ is:
$$
(\overline{a}a\overline{a}b, 2, a\overline{a}ba\overline{a}\overline{b}a\overline{a}, -1,
a\overline{a}\overline{b}a\overline{a}\overline{b}a\overline{a},0,
a\overline{a}\overline{b}a\overline{a}).
$$
\end{example}

Second case: We assume that 
$\overline{p}$
is a factor of $p^2$. Then we can write $p =rs$ with 
$\overline{p} = sr$ and $r = \overline{r}$, $s = \overline{s}$.
We allow $ r =1$, hence the second case includes the case $p =
\overline{p}$. In fact, if $ r =1$, then below  we obtain 
the usual definition of  $p$-stable normal
  form. 
Moreover, by switching to some conjugated word of $p$
we could always assume that $r \in \{1, a\}$ for some letter $a$
being fixed by the involution, $a = \overline{a}$,
but this switch  is not made here.
The idea is to replace each maximal factor of the form $(rs)^{\alpha}r$
with $\alpha \geq 2$ by a sequence $rs,\alpha - 2,sr$.
In this notation 
$\alpha - 2$ is representing the factor $(rs)^{\alpha -2}r
=p^{\alpha - 2}r = r \overline{p}^{\alpha - 2} $.

The {\em $p$-stable normal
  form} (second kind) of  $w \in A^*$
is now a
shortest sequence ($k$ is minimal)
\[
( u_{0},\alpha _{1},u_{1},\ldots ,
\alpha _{k},u_{k}) 
\]
such that $k\geq 0$, $u_{0},u_{i}\in A^{*},\;\alpha _{i}\geq 0$ for
$1\leq i\leq k$, and the following conditions are satisfied:
\begin{itemize}
\item  $w=u_{0}p^{\alpha _{1}} r u_{1}\cdots 
p^{\alpha _{k}} r u_{k}$.

\item  $k=0$ if and only if $p^2 r$ is 
not  a  factor of $w.$

\item  If $k\geq 1$, then: 
\begin{eqnarray*} 
u_{0} &\in  &A^{*}rs\setminus (A^{*}p^{2}rA^{*} \cup A^*rs rs), \\
u_{i} &\in  &( A^{*}rs \cap srA^{*}) 
\setminus (srsr A^* \cup A^{*}p^{2}rA^{*}\cup A^*rs rs)\,
\mbox{ for }1\leq i<k, \\
u_{k} &\in & srA^{*}\setminus (A^{*}p^{2}rA^{*}\cup srsrA^*).
\end{eqnarray*}
\end{itemize}

Since  $\overline{rs} =sr$,
the $p$-stable normal form of $\overline{w}$
becomes
$$( \overline{u_{k}},\alpha _{k},u_{1},\ldots 
,\alpha _{1},\overline{u_{0}}). $$ So, for the second kind no negative 
integers interfere.

\begin{example} Let $ p = a\overline{a}b$ with $b =\overline{b}$.
Then $r =   a\overline{a}$ and $s= b$. Let 
$w = \overline{a}p^4ap^3 a$
Then the $p$-stable normal form of $w$ is:
$$
(\overline{a}ba\overline{a}b, 2, ba\overline{a}baa\overline{a}b,0,ba\overline{a}ba).$$
\end{example}

In both cases 
we can write the $p$-stable normal form of $w$
as a sequence $$( u_{0},\alpha _{1},u_{1},\ldots ,\alpha _{k},u_{k}) $$
where $u_i$ are words and $\alpha _{i}$ are integers.

For every finite semigroup $S$ there is a number $c(S)$ such that for
all $s\in S$ the element $s^{c(S)}$ is idempotent, i.e., $s^{c(S)} =
s^{2c(S)}$.  It is clear that the number $c(M)$ for our monoid $M
\subseteq \mathbb{B}^{2n\times 2n}$ is the same as the number
$c(\mathbb{B}^{n\times n})$.  It is well-known \cite{mar77} that we
can take $c(\mathbb{B}^{n\times n}) = n!$
 (it is however more convenient to define $c(M) = 3$ for
$n = 1$). Hence in the following
$c(M) = \max\{3,n!\}$.  

For specific situations this
might be an overestimation, but
this choice  guarantees  $h(uv^{c(M)}w) =
h(uv^{2c(M)}w)$ for all $u,v,w\in \Gamma^*$ and all $h:\Gamma^* \rightarrow M$.

Now, let $w,w'\in\Gamma^*$ be words such that the $p$-stable normal
forms are identical up to one position where for $w$ appears an
integer $\alpha_i$ and for $w'$ appears an integer $\alpha_i'$.
We know $h(w) = h(w')$ whenever
 the following conditions are satisfied: $\alpha_i \cdot
\alpha_i' > 0$, $|\alpha_i| \geq c(M)$, $|\alpha_i'| \geq c(M)$, and
$\alpha_i \equiv \alpha_i' \pmod {c(M)}$.  Then we have $h(w) = h(w')$.
This is the reason to change the syntax of the $p$-stable normal form.
Each non-zero integer $\alpha'$ is written as $\alpha' = \varepsilon (q +
\alpha c(M))$ where $\varepsilon, q, \alpha$ are uniquely defined by
$\varepsilon \in \{ +1, -1 \}$, $0\leq q < c(M)$, and $\alpha \geq 0$.
For $\alpha'= 0 $ we may choose $\varepsilon = q = \alpha = 0$.
We shall read $\alpha$ as a variable ranging over
non-negative integers, but $\varepsilon$, $q$, and $c(M)$  
are viewed as constants. In fact, if $|\alpha'| < c(M)$,
then we best view $\alpha$ also as a constant in order to avoid
problems with the constraints.

Let $u$, $v$, and $w$ be words such that $uv= w$ holds.
Write these words in  their $p$-stable normal
forms: 
\[
\begin{array}{ll}
u \colon &(u_{0},\varepsilon_{1}(q_{1}+\alpha _{1}c(S)),
u_{1},\ldots ,\varepsilon_{k}(q_{k}+\alpha
_{k}c(S)),u_{k}), \\
v \colon &(v_{0},\varepsilon'_{1}(s_{1}+\beta _{1}c(S)),
v_{1},\ldots ,\varepsilon'_{\ell}(s_{\ell }+\beta
_{\ell }c(S)),v_{\ell }), \\
w  \colon &(w_{0},\varepsilon''_{1}(t_{1}+\gamma _{1}c(S)),
w_{1},\ldots ,\varepsilon''_{m}(t_{m}+\gamma
_{m}c(S)),w_{m}).
\end{array}
\]

Since $uv = w$ there are many identities. For example, for $k,\ell \geq 2$
we have $u_{0}=w_{0}$, $v_{l}=w_{m}$, $%
q_{1}=t_{1}$, $\alpha _{1}=\gamma _{1}$, etc.  What exactly happens
depends only on the $p$-stable normal form of the product
$u_{k}v_{0}$.  There are several cases, which easily can be listed.
We treat only one of them, which is in some sense the worst case in
order to produce a large exponent of periodicity.  This is the case
where $ p = rs$ with $r = \overline{r}$ and $s = \overline{s}$.  Then
it might be that $u_k= srsr_1 $ and $v_0 = r_2srs$ with $r_1r_2 =
r$ (and $r_1 \not= 1 \not= r_2$).
Hence we have  $u_{k}v_{0}=sp^{3}$ and $k+\ell =m+1$.
It follows $\alpha_1 = \gamma_1, \ldots, \alpha_{k-1} = \gamma_{k-1}$,
$\beta_2 = \gamma_{k+1}, \ldots, \beta_{\ell} = \gamma_{m}$,
and there is only one non-trivial identity:

\[q_{k}+s_{1}+4 +(
\alpha _{k}+\beta _{1})c(S)=t_{k}+\gamma _{k}c(S).
\]
Since by assumption $c(S)\geq 3$, the case $u_{k}v_{0}=sp^{3}$
leads to the identity:
\[
\gamma _{k}= \alpha _{k}+\beta _{1}
+ c\mbox{ with } c \in  \{0,1,2\}.
\]

Assume now that $\alpha _{k}\geq 1$ and $ \beta _{1}\geq 1$.
If we replace $\alpha _{k}$, $\beta _{1}$, and $\gamma _{k}$
by some $\alpha _{k}' \geq 1$, $\beta _{1}' \geq 1$, and $\gamma _{k}'
\geq 1$
such that still $\gamma'_{k}= \alpha'_{k}+\beta'_{1}
+ c$, then we obtain new words $u'$, $v'$, and $w'$ with the same
images under $h$ in $M$ and still the identity $u'v'=w'$.

What follows then is completely analogous to what has been done in
detail in \cite{kp96,gut2000stoc,hagenahdiss2000,die98lothaire}.
Using the $p$-stable normal form we can associate with an equation $L
= R$ of denotational length $d$ together with its solution $\sigma:
\Omega \rightarrow \Gamma^*$ some linear Diophantine system of
$d$ equations in at most
$3d$ variables. The variables range over natural numbers since zeros are
substituted.  (In fact the 
number of variables can be reduced to be at most
$2|\Omega|$). The parameters of this system are such that maximal
size of a minimal solution (with respect to the component wise partial
order of $\mathbb{N}^d$) is in $\OO(2^{1.6d})$ with the same approach
as in
\cite{kp96}. This tight bound is based in turn on the work of \cite{gs78}; a
more moderate bound $2^{\OO(d)}$ (which is enough for our purposes) is
easier to obtain, see e.g.\ \cite{die98lothaire}.  The maximal size of
a minimal solution of the linear Diophantine system has a backward
translation to a bound on the exponent of periodicity.  For this
translation we have to multiply with the factor $c(M) \in 2^{\OO(n\log
  n)}$ and to add $c(M) +1$.  Putting everything together we 
obtain the claim of the  proposition.
\qed
\end{proof}

\section{Exponential Expressions}

During the procedure which solves 
Theorem~\ref{theo3}  various other equations with constraints are considered
but the monoid $M$ will not change.

There will be not enough space to write down the equation $L=R$ in
plain form, in general. In fact, there is a provable exponential lower
  bound for the length $|LR|$ in the worst case which we can meet
  during the procedure.  In order to overcome this difficulty
Plandowski's method uses data compression for words in $(\Gamma \cup
\Omega)^*$ in terms of exponential expressions.

Exponential expressions (their evaluation and their size) are inductively
defined:
\begin{itemize}
\item Every word $w \in \Gamma^*$ denotes an exponential expression.
      The evaluation $\eval(w)$ is equal to $w$, its size $\| w \|$ is
      equal to the length $|w|$.
\item Let $e$, $e'$ be exponential expressions. Then $e e'$ is an
      exponential expression. Its evaluation is the concatenation
      $\eval(e e')=\eval(e) \eval(e')$, its size is
      $\| e e' \| = \| e \| + \| e' \|$.
\item Let $e$ be an exponential expression and $k \in \N$. Then $(e)^k$
      is an exponential expression. Its evaluation is
      $\eval((e)^k) = (\eval(e))^k$, its size is
      $\| (e)^k \| = \log(k)+ \| e \|$ where $\log(k) = \max \{1, \lceil \log_2(k) \rceil\}$.
\end{itemize}
It is not difficult to show that the
length of $\eval(e)$ is at most exponential in the size of $e$,
a fact which is, strictly speaking, not needed for the proof of
Theorem~\ref{theo3}. What we need  however is 
the next lemma. Its proof can be done easily by structural 
induction and it is omitted.

\begin{lemma} \label{expfac}
Let $u \in \Gamma^*$ be a factor of a word $w\in \Gamma^*$. Assume
that $w$ can be represented by some exponential expression of size $p$.
Then we find an exponential expression of size at most $p^2$ that
represents $u$.
\end{lemma}

 We say that an exponential expression $e$ is {\em admissible\/}, if
its size $\| e \| $ is bounded by some fixed polynomial in the input
size of $E_0$. The lemma above states that if $e$ is admissible, then
we find admissible exponential expressions for all factors of
$\eval(e)$. But now the admissibility is defined with respect to some
polynomial which is the square of the original polynomial, so, in a
nested way, we can apply this procedure a constant number of times,
only.  In our application the nested depth does not go beyond two.

The next lemma is straightforward
since we allow a polynomial space bound without any
time restriction. Again, the proof is left to the reader.

\begin{lemma} \label{basiceval}
The following two problems can be
solved in $\PSPACE$.
\smallbreak
INPUT: Exponential expressions  $e$ and $e'$.

QUESTION: Do we have $\eval(e)=\eval(e')$?

\smallbreak
INPUT: A  mapping
$h: \Gamma \to M$ and an exponential expression $e$.

OUTPUT: The matrix $h(\eval(e)) \in M$.
\end{lemma}

\begin{remark}
  \label{polyeval}%
The computation
 above can actually be performed in polynomial
time, but this is not evident for the first question, see \cite{ESA::Plandowski1994} for details.
\end{remark}

Henceforth we allow that the part $L=R$ of an equation with constraints
may also be given by a pair of exponential expressions $(e_L, e_R)$ with
$\eval(e_L)=L$ and $\eval(e_R)=R$.   
 We say that $E = (\Gamma, h , \Omega, \rho ; e_L = e_R)$
is {\em admissible\/}, if $e_L e_R$
is admissible,  
$|\Gamma \setminus \Gamma_0|$ has polynomial size,
$\Omega \subseteq \Omega_0$, and 
$h(a)=h_0(a)$ for $a \in \Gamma \cap \Gamma_0$.

 For two admissible equations with constraints
$E = (\Gamma, h , \Omega, \rho ; e_L=e_R)$
and $E'= (\Gamma, h , \Omega, \rho ; e'_L= e'_R)$ 
we write $E \equiv E'$,
if $\eval(e_L)=\eval(e'_L)$
and $\eval(e_R)=\eval(e'_R)$
 as strings in $(\Gamma \cup \Omega)^*$. This means that they 
represent exactly the same equations.

\section{Base Changes}
In this section we fix a mapping $h: \Gamma \to M$ which respects
the involution.
Let $(\Gamma',\invol)$ be an alphabet with involution 
and let $\beta: \Gamma' \to \Gamma^*$ be some 
mapping $\beta$ such that 
$\beta(\overline{a}) = \overline{\beta(a)}$ for all $a \in \Gamma'$.
We define  $h' : \Gamma' \to M $ such that $h'  = h \beta $.
We also extend to a homomorphism
$\beta: (\Gamma' \cup \Omega)^* \to (\Gamma \cup \Omega)^*$  by leaving
the variables invariant. 

 Let 
$E'=(\Gamma', h' , \Omega, \rho ; L'=R').$
be an equation with constraints. The
{\em base change\/} $\beta_*(E')$ is defined by 
$$\beta_*(E') = (\Gamma, h, \Omega, \rho ; \beta(L')=\beta(R')).$$

We also refer to $\beta: \Gamma' \to \Gamma^*$ as a base change and we
say that $\beta$ is {\em admissible\/}, if $|\Gamma'| $ has polynomial
size and if $\beta(a)$ can be represented by some admissible
exponential expression for all $a \in \Gamma'$.

\begin{remark}
  \label{basepoly}%
  If $\beta: \Gamma' \to \Gamma^*$ is an admissible base change and if
  $L'=R'$ is given by a pair of admissible exponential expressions,
  then we can represent $\beta_*(E')$ by some admissible equation with
  constraints. A representation of $\beta_*(E')$ is computable in
  polynomial time.
\end{remark}

\begin{lemma} \label{basechange}
Let $E'$ be an equation with constraints  and
$\beta: \Gamma' \to \Gamma^*$ be a base change. If
$\sigma'$ is a solution of $E'$, then
$\sigma=\beta \sigma'$ is a solution of $\beta_*(E')$.
\end{lemma}

\begin{proof}
Clearly $\sigma(\overline{X})=\overline{\sigma(X)}$ and
$h\sigma(X)=h\beta\sigma'(X)=h'\sigma'(X)=\rho(X)$ for all $X \in \Omega$.
Next by definition $\sigma(a)=a$ for $a\in \Gamma$ and $\beta(X)=X$ for
$X \in \Omega$. Hence $\sigma\beta(a)=\beta\sigma'(a)$ for $a \in \Gamma'$ and
therefore $\sigma\beta=\beta\sigma': (\Gamma' \cup \Omega)^* \to \Gamma^*$.
This means $\sigma\beta(L)=\beta\sigma'(L)=\beta\sigma'(R)=\sigma\beta(R)$ since
$\sigma'(L)=\sigma'(R)$.\qed
\end{proof}

The lemma above leads to the first rule.\\

{\bf Rule~1} {\em If $E$ is of the form $\beta_*(E')$
and if we are  looking for a solution of $E$, then 
it is enough to find a solution for $E'$. Hence, during 
a non-deterministic search
we may replace 
$E$ by $E'$.\\}

\begin{example}\label{ex-beta}
 Consider the following equation $E$ with constraints over 
$\Gamma = \{a,b,c, \oa, \ob, \oc \}$:
\[
       X \ox 
	= Y \ob \oc \ob \oa \ob \oc \ob Y Z a b c b \oy.
\]
Let there be the constraints for $X$ and $Z$ saying 
 $X \in \Gamma^{300}\Gamma^*$ and  $Z \in \ob \oc \ob \oa \Gamma^* $.
 Define $\Gamma' = \{a,b,\oa,\ob\}$ and a base change 
$\beta: \Gamma' \to \Gamma^*$ 
by $\beta(a)=abcb$ and $\beta(b)=bcb$.
Then the equation $E$ is of the form $\beta_*(E')$ where $E'$
is given by 
\[
        X \ox  = Y \oa \ob Y Z a \oy
\]
 and the new (and sharper) constraint
for $Z$ is simply $Z \in \oa{\Gamma'}^*$, 
for $X$ we may sharpen the constraint to 
$X \in {\Gamma'}^{100}{\Gamma'}^*$
 According to Rule~1 it is enough to solve $E'$.
The effect of the base change 
$\beta$ is that the equation $E'$ is shorter and the 
alphabet of constants becomes smaller, since the letter $c$ is not
used anymore. Note also that the length restriction 
on  $X$ became smaller, too.
However this has a prize; in general, $E=\beta_*({E'})$ might
have a solution, whereas $E'$ is unsolvable. As we will see later,
our guess has been correct in the sense that $E'$ still has a solution.
\end{example}

\section{Projections}

Let $(\Gamma,\invol)$ and $(\Gamma',\invol)$ be alphabets with involution
such that $(\Gamma,\invol) \subseteq (\Gamma',\invol)$.  A
{\em projection\/} is a homomorphism $\pi:{{\Gamma'}}^* \to
\Gamma^*$ such that both $\pi(a) = a$ for $a\in\Gamma$ and
$\pi(\overline{a}) = \overline{\pi(a)}$ for all $a\in{\Gamma'}$.
If $h:\Gamma\to M$ is given, then a projection $\pi$
defines also $h':{\Gamma'}\to M$ by $h' = h\pi$.

Let $E$ be an equation with constraints
$E = (\Gamma , h , \Omega, \rho; L=R).$
Then we can define an equation with constraints $\pi^*(E)$ by
$$\pi^*(E) = ({\Gamma'} , h\pi , \Omega , \rho ; L=R).$$
  The difference between $E$ and $\pi^*(E)$ is only in the
alphabets of constants and in the mappings $h$ and $h' = h\pi$.
Note that every projection $\pi: {\Gamma'}^*\to \Gamma^*$
defines a base change $\pi_*$ such that $\pi_*\pi^*(E) = E$.

\begin{lemma}
\label{projchar}%
Let $E=  (\Gamma , h , \Omega , \rho ; L=R)$ 
and $E' = ({\Gamma'} , h' , \Omega , \rho; L=R)$
be equations with constraints.
Then the following two statements hold.
  \begin{enumerate}
    \renewcommand{\labelenumi}{\roman{enumi})}
  \item There is a projection $\pi: {\Gamma'}^* \to \Gamma^*$ such that
    $\pi^*(E) = E'$, if and only if both $h'({\Gamma'}) \subseteq
    h(\Gamma^*)$ and for all $a\in{\Gamma'}$ with $a=\overline{a}$ 
    there is some $w \in \Gamma^*$ with  $w = \overline{w}$ 
    such that  $h'(a) = h(w)$.
  \item If we have $\pi^*(E) = E'$ and if $\sigma':\Omega \to
    {\Gamma'}^*$ is a solution of $E'$, then we effectively find a
    solution $\sigma$ for $E$ such that
    $|\sigma(L)| \leq 2|M||\sigma'(L)|$.
  \end{enumerate}
\end{lemma}

\begin{proof}
  i) Clearly, the only-if condition is satisfied by the definition of
  a projection since then $h' = h\pi$.  For the converse, assume that
  $h'({\Gamma'}) \subseteq h(\Gamma^*)$ and that $a = \overline{a}$ implies
  $h'(a) \in h(\{ w\in\Gamma^* \mid w=\overline{w} \} )$.  Then for
  each $a \in {\Gamma'}\setminus\Gamma$ we can choose a word
  $w_a\in\Gamma^*$ such that $h'(a) = h(w_a)$.  We can make the choice
  such that $w_{\overline{a}} = \overline{w_a}$ for all $a \in
  {\Gamma'}\setminus\Gamma$.  If $a\neq \overline{a}$, then we can find
  $w_a$ such that $|w_a| < |M|$, since we
  can take the shortest word $w_a\in \Gamma^*$ such that $h(w_a) = h'(a)
  \in M$.  For $a=\overline{a}$ we know that there is some word
  $w_a\in\Gamma^*$ with $h'(a) = h(w_a)$ and $w_a=\overline{w_a}$.  Hence we
  can write $w_a = vb\overline{v}$ with $b\in\Gamma\cup \{1\}$ and
  $b=\overline{b}$.  For $b\neq 1$ we can demand $|w_a| \leq 2|M|-1$.
  For $b=1$ we can demand $|w_a| \leq 2|M|-2$.  Thus, we find a projection
  $\pi:{\Gamma'}^* \to \Gamma^*$ such that $\pi^*(E) = E'$ and
  moreover, $|\pi(a)| < 2|M|$ for all $a\in{\Gamma'}$.

  ii) Using the reasoning in the proof of i) we may assume that
  $\pi:{\Gamma'}^* \to \Gamma^*$ satisfies $|\pi(a)| < 2|M|$
  for all $a\in{\Gamma'}$.  Since $\pi$ defines a base change with
  $\pi_*(E') = E$, we know by Lemma~\ref{basechange} 
  that $\sigma = \pi\sigma'$ is a solution of $E$.  
  Clearly, $|\sigma(L)| = |\pi\sigma'(L)| \leq 2 |M| |\sigma'(L)|$. \qed
\end{proof}

\begin{remark}\label{decideproj}
  In the following we will meet the problem to decide whether there
  is a projection $\pi:{\Gamma'}^* \to \Gamma^*$ such that
  $\pi^*(E) = E'$.  We actually need not too much  space for this
  test.
It is not necessary to write down $\pi$. We
  can use the criterion  in the lemma above and Lemma~\ref{basicops}.
  Then we have to store
  in the working space only some Boolean matrices of $\B^{2n\times
    2n}$.  In particular, if $n$ is a constant (or logarithmically
  bounded in the input size), then the test $\exists\pi: \pi^*(E) = E'$ can be
  done in polynomial time.  However, if $n$ becomes a substantial part
  of the input size, then the test might be difficult in the sense
  that we might need the full power of $\PSPACE$.
\end{remark}

The lemma above leads now to the second  rule.\\

{\bf Rule~2} {\em If $\pi$ is a projection 
and if we are  looking for a solution of $E$, then 
it is enough to find a solution for $\pi^*(E)$. Hence, during 
a non-deterministic search
we may replace 
$E$ by $\pi^*(E)$.\\}

\begin{example}\label{ex-pi}
Let us continue with the equation which has been obtained by the
transformation in Example~\ref{ex-beta}. 
In order to simplify notations, we will call $E$ the equation
$	 X \ox  = Y \oa \ob Y Z a \oy$,
and $\Gamma = \{ a,b,\oa, \ob \}$.

Remember that the constraint
on $X$ demanded a rather long 
solution.  Therefore we may reintroduce a letter $c$
and put $\Gamma' = \{a,b,c,\oa, \ob, \oc \}$. Then we may define
a projection $\pi: \Gamma' \to \Gamma^*$ by, say, 
$\pi(c)=b^{100}$. The equation $E' = \pi^*(E)$ looks as above,
but in $E'$ we may change the constraint for $X$. We may sharpen the new
constraint for $X$ to be $X \in \Gamma^* c \Gamma^*$.
Thus, the solution for $X$ might be very short now.
\end{example}

\section{Partial Solutions}

Let $\Omega' \subseteq \Omega$ be a subset of the variables which is closed
under involution.  We assume that there  is a mapping $\rho': \Omega'
\to M$ with $\rho'(\overline{x}) = \overline{\rho'(x)}$, but
we do not require that $\rho'$ is the restriction of $\rho: \Omega
\to M$.  Consider an equation with constraints
$E = (\Gamma , h , \Omega,\rho; L=R).$
A {\em partial solution} is a mapping $\delta:\Omega \to \Gamma^*
\Omega' \Gamma^* \cup \Gamma^*$ such that the following conditions are
satisfied:
\begin{enumerate}
  \renewcommand{\labelenumi}{\roman{enumi})}
\item \parbox{3cm}{$\delta(X) \in \Gamma^* X \Gamma^*$} for all
  $X\in\Omega'$,
\item \parbox{3cm}{$\delta(X) \in \Gamma^*$} for all
  $X\in\Omega\setminus \Omega'$,
\item \parbox{3cm}{$\delta(\overline{X}) = \overline{\delta(X)}$}
  for all $X\in\Omega$.
\end{enumerate}
The mapping $\delta$ is extended to a homomorphism $\delta:(\Gamma
\cup \Omega)^* \to (\Gamma \cup \Omega')^*$ by leaving the
elements of $\Gamma$ invariant.  Let
$ E' = (\Gamma , h , \Omega',\rho'; L'=R')$ 
be another equation with constraints
(using the same $\Gamma$ and $h$).  We write $E'
= \delta_*(E)$, if there exists some partial solution $\delta:\Omega
\to \Gamma^* \Omega \Gamma^* \cup \Gamma^*$ such that the following
conditions hold: 
$L' = \delta(L)$, $R' = \delta(R)$, $\rho(X) = h(u)\rho'(X)h(v)$ for
$\delta(X) = uXv$, and $\rho(X) = h(w)$ for $\delta(X) =
w\in\Gamma^*$.

\begin{lemma}
  \label{shiftsol}%
  In the notation of above, let $E' = \delta_*(E)$ for some partial solution
  $\delta: \Omega \to \Gamma^* \Omega \Gamma^* \cup \Gamma^*$.
  If $\sigma'$ is a solution of $E'$,
  then $\sigma = \sigma'\delta $ is a
  solution of $E$.  Moreover, we have $\sigma(L) = \sigma'(L')$
  and $\sigma(R) = \sigma'(R')$.

\end{lemma}

\begin{proof}
  By definition, $\delta$ and $\sigma'$ are extended to homomorphisms
  $\delta: (\Gamma\cup\Omega)^* \to (\Gamma\cup\Omega')^*$ and
  $\sigma': (\Gamma \cup \Omega')^* \to \Gamma^*$ leaving the
  letters of $\Gamma$ invariant.  Since $E' = \delta_*(E)$ we have
  $\delta(L) = L'$ and $\delta(R) = R'$.  Since $\sigma'$ is a
  solution, we have $\sigma(L) = \sigma'\delta(L) = \sigma'(L') =
  \sigma'(R') = \sigma'\delta(R) = \sigma(R)$ and $\sigma$ leaves the
  letters of $\Gamma$ invariant.  The solution $\sigma'$ satisfies
  $h\sigma'(X) = \rho'(X)$ for all $X\in\Omega'$.  Hence, if $\delta(X)
  = uXv$, then $\rho(X) = h(u)\rho'(X)h(v) = h(u\sigma'(X)v) =
  h\sigma'(uXv) = h\sigma'\delta(X) = h\sigma(X)$.  If $\delta(X) =
  w\in\Gamma^*$, then $\sigma(X) = \sigma'\delta(X) = w$ and $\rho(X)
  = h(w)$, again by the definition of a partial solution.
  \qed
\end{proof}

\begin{lemma}
  \label{shifttest}%
  The following problem can be solved in $\PSPACE$.
\smallbreak
  INPUT: Two equations with constraints 
   $E = (\Gamma , h , \Omega,\rho; e_L = e_R)$ 
and $E' = (\Gamma , h , \Omega', \rho'; e_{L'} = e_{R'})$.

  QUESTION: Is there some partial solution $\delta$ 
	such that $\delta_*(E) \equiv E'$?

\smallbreak
  Moreover, if $\delta_*(E) \equiv E'$ is true, then there are
  admissible exponential expressions $e_u$, $e_v$ for each $X\in\Omega'$
  and an admissible exponential expression $e_w$ for each $X \in
  \Omega\setminus\Omega'$ such that
$$\begin{array}{rcll}
    \delta(X) & = & \eval(e_u)X\eval(e_v) &\quad \mathrm{ for }\; X\in\Omega', \\
    \delta(X) & = & \eval(e_w) & \quad \mathrm{ for }\; X\in\Omega\setminus\Omega'.
\end{array}
$$
\end{lemma}

\begin{proof}
  Let $L = \eval(e_L)$, $R = \eval(e_R)$, $L' = \eval(e_{L'})$, and
  $R' = \eval(e_{R'})$. The non-deterministic algorithm works as follows:
  
  For each
  $X\in\Omega'$ we guess admissible exponential expressions $e_u$ and
  $e_v$ with $\eval(e_u),\eval(e_v) \in \Gamma^*$.  We define an
  exponential expressions $e_X = e_uXe_v$ and $\delta(X) =
  \eval(e_X)$. For each $X \in \Omega\setminus\Omega'$ we guess an admissible
  exponential $e_X$ with $\eval(e_X) \in \Gamma^*$ and $\delta(X) =
  \eval(e_X)$. 
  
  Next we verify whether or not $\delta_*(E) \equiv E'$.
  During this test we have to create an exponential expression $f_L$
  (and $f_R$, resp.) by replacing $X$ in $e_L$ (and $e_R$, resp.) with
  the expression $e_X$.  This increases the size in the worst case 
  by a factor of $\max \{ ||e_X|| \mid X\in\Omega \}$.
The other tests whether
$\rho(X) = h(u)\rho'(X)h(v)$ for $\delta(X) = uXv$ and $\rho(X) =
h(w)$ for $\delta(X) = w\in\Gamma^*$ involve admissible exponential
expressions over Boolean matrices and can be done in polynomial time.

  The correctness of the algorithm follows from our general assumption
  that all $X\in\Omega$ appear in $LR\overline{LR}$.  Therefore, if we
  have $\delta_*(E) \equiv E'$, then $\delta(X)$ (or
  $\delta(\overline{X})$) appears necessarily as a factor in $L'R' = \delta(LR)$.
    Hence $\delta(X)$ has an
  exponential expression of polynomial size by Lemma~\ref{expfac}.
  Therefore guesses of $e_u$, $e_v$, and $e_w$ as above are possible
without running out of space. \qed
\end{proof}

\begin{remark}\label{npeval}
Actually, the test for
  $\delta_*(E) \equiv E'$ can be performed in non-deterministic 
polynomial time by Remark~\ref{polyeval}.
\end{remark}

The lemma above leads to the third and last  rule.\\

{\bf Rule~3} {\em If $\delta$ is a partial solution 
and if we are  looking for a solution of $E$, then 
it is enough to find a solution for $\delta_*(E)$. Hence, during 
a non-deterministic search
we may replace 
$E$ by $\delta_*(E)$.\\}

\begin{remark}\label{projtest}
  We can think of a partial solution $\delta:\Omega \to \Gamma^*
  \Omega' \Gamma^* \cup \Gamma^*$ in the following sense.  Assume we
  have an idea about $\sigma(X)$ for some $X\in\Omega$.  Then we might
  guess $\sigma(X)$ entirely.  In this case we can define $\delta(X) =
  \sigma(X)$ and we have $X\not\in\Omega'$.  For some other $X$ we
  might guess only some prefix $u$ and some suffix $v$ of $\sigma(X)$.
  Then we define $\delta(X) = uXv$ and we have to guess some
  $\rho'(X)\in M$ such that $\rho(x): h(u)\rho'(X)h(v)$.  If our guess
  was correct, then such a matrix $\rho'(X)\in M$ must exist.  We have
  partially specified the solution and applying Rule~3, we continue
  this process by replacing the equation $L=R$ by the new equation
  $\delta(L) = \delta(R)$.
\end{remark}

\begin{example}\label{ex-delta}
We continue with our running example.
After renaming, the equation $E$ is given by
\[
	 X \ox  = Y \oa  \ob  Y Z a \oy,
\]
and 
the alphabet of constant is given by $\Gamma = \{ a, b,c, \oa, \ob, \oc \}$.
The constraints are $X\in \Gamma^* c \Gamma^*$
and  $Z \in \oa \{a,b,\oa, \ob \}^*$.

 We may guess the partial solution as follows:
$\delta(X)=aX$, $\delta(Y)=Y$, and $\delta(Z)=\oa b$.
The new equation $\delta_*(E)$ is
\[
     a X \ox \oa = Y \oa  \ob Y \oa b  a \oy.
\]
The remaining constraint is that the solution for $X$ has to use 
the letter $c$.

 The process can continue, for example, 
we can apply Rule~1 again by defining
another  base change $\beta(b) = ba$ to get the equation
\[
   a X \ox \oa = Y \ob Y \oa b \oy
\]
 over $\Gamma = \{ a,b,c,\oa, \ob, \oc\}$.
 Since the last equation has a solution 
(e.g., given by $\sigma(X)=b c \oc \ob \ob a b c$ and 
$\sigma(Y)=a b c\oc \ob$), 
the first equation with constraints in Example~\ref{ex-beta}
has a solution too.
\end{example}

\section{The Search Graph and Plandowski's Algorithm}

 In the following we show that there is some fixed polynomial 
(which can be calculated from the presentation below)  such that
the high-level description of Plandowski's algorithm is as follows:
On input $E_0$ compute the maximal space bound, given by the polynomial,
to be used by the procedure. Then apply non-deterministically 
Rules~1, 2, and 3 until an equation with a trivial solution is found.

 From the description above it follows that the 
specification of the algorithm just uses Rules~1, 2, 3. 
The algorithm is simple but it demands a good heuristics to 
explore the search graph.
The hard part is to prove that this schema is correct;
for this we have to be more precise.

The {\em search graph} is a directed graph: 
The nodes are admissible equations with constraints.
For two nodes $E$, $E'$, we define an arc $E \rightarrow E'$, if
there are an admissible base change $\beta$, a
projection $\pi$, and a partial solution $\delta$ 
such that $\delta_*(\pi^*(E)) \equiv \beta_*(E')$.

\begin{lemma} \label{searchgrapharc}
The following problem can be decided in $\PSPACE$.

INPUT: Admissible equations with constraints $E$ and $E'$.

QUESTION: Is there an arc $E \rightarrow E'$ in the search graph?
\end{lemma}

\begin{proof}
We first guess some alphabet
$({\Gamma'}',\invol)$  of polynomial size
together with $h'':{\Gamma'}' \to M$. Then we guess some
 admissible base change
$\beta: {\Gamma'} \to {\Gamma''}^*$ such that $h'= h''\beta$
and we compute $\beta_*(E')$.

Next we guess some admissible equation with constraints $E''$ 
which uses ${\Gamma''}$  and $\Omega$.
We 
check using  Lemma~\ref{shifttest} that there is
some  partial solution $\delta: \Omega \to
{\Gamma''}^* \Omega' {\Gamma''}^* \cup {\Gamma''}^*$
such that 
$\delta_*(E'') \equiv \beta_*(E')$.
(Note that every
equation with constraints $E''$ 
satisfying $\delta_*(E'') \equiv \beta_*(E')$
for some $\delta$ is admissible by Lemma~\ref{expfac}.)
Finally we 
check using Remark~\ref{projtest}
and  that there is some  projection
$ \pi:{\Gamma''} \to \Gamma$ 
such that $\pi^*(E) \equiv E''$.
We obtain
$\delta_*(\pi^*(E)) \equiv \beta_*(E')$.\qed
\end{proof}

\begin{remark}\label{arctest}
Following Remarks~\ref{polyeval} and~\ref{npeval} the problem in
Lemma~\ref{searchgrapharc} can be decided in non-deterministic 
polynomial time, if the monoid $M$ is not part of the input and
viewed as a constant. If, as in our setting,
 $M$ is part of the input, then $\PSPACE$ is the best we can prove,
because the test for the projection becomes difficult.
\end{remark}

Plandowski's algorithm works as follows:

\medbreak\nobreak{
{\bf begin}

\atab$E:=E_0$

\atab{\bf while} $\Omega \neq \emptyset$ {\bf do}

\atab\atab Guess an admissible equation $E'$ with constraints 

\atab\atab Verify that $E \rightarrow E'$ is an arc in the search graph

\atab\atab$E:=E'$

\atab{\bf endwhile}

\atab{\bf return} ``$\eval(e_L)=\eval(e_R)$''

{\bf end}
} \medbreak

By Rules~1--3 (Lemmata~\ref{basechange}, \ref{projchar}~$ii)$,  and \ref{shiftsol}),
 if $E \rightarrow E'$ is an arc
in the search graph and $E'$ is solvable, then $E$ is solvable, too.
 Thus, if the algorithm returns {\em true\/}, then $E_0$ is
solvable. The proof of Theorem~\ref{theo3}
is therefore reduced to the statement that if $E_0$
is solvable, then the search graph contains a path to some
node without variables and the exponential expressions defining
the equation evaluate to the same word. This existence proof
is the hard part, it
covers the rest of the paper.

\begin{remark}
  \label{doublesize}%
If $E \rightarrow E'$ is due to some   $\pi: {\Gamma''}^* \to
\Gamma^*$,
$\delta: \Omega \to
{\Gamma''}^* \Omega' {\Gamma''}^* \cup {\Gamma''}^*$,
 and
$\beta: {\Gamma'}^* \to {\Gamma''}^*$, then a solution
$\sigma': \Omega' \to {\Gamma'}^*$ of $E'$ yields the solution
$\sigma = \pi(\beta\sigma')\delta$. Hence we may assume that
the length of a solution has
increased by at most an exponential factor by Lemma~\ref{projchar} $ii)$. Since we are going
to perform the search in a graph of at most exponential size,
we get automatically a doubly exponential upper bound for the length
of a minimal solution by backwards computation on such a path.
 This is still the best known upper
bound (although an singly exponential bound is conjectured), see
\cite{pla99stoc}.
\end{remark}

\section{Free Intervals}

In this section we introduce the notion of {\em free interval} in order
to cope with long factors in the solution which are not related to
any cut. If there were no constraints, then these factors would not appear
in a minimal solution. In our setting we cannot avoid these factors.

For a word $w \in \Gamma^*$ we let $\{0, \ldots, |w|\}$ be the set of its
{\em positions\/}. The interpretation is that factors of $w$ are between
positions. To be more specific let $w=a_1 \cdots a_m$,
$a_i \in \Gamma$ for $1 \leq i \leq m$. Then $[\alpha, \beta]$ with
$0 \leq \alpha < \beta \leq m$ is called a \emph{positive interval}
and the factor $w[\alpha,\beta]$ is defined by the word
$w[\alpha,\beta]= a_{\alpha+1} \cdots a_\beta$.

It is convenient to have an involution on the set of intervals. Therefore
$[\beta, \alpha]$ is also called an interval (but it is never
positive),
 and we define
$w[\beta,\alpha]=\overline{w[\alpha,\beta]}$.
We allow also $\alpha = \beta$ and we define $w[\alpha,\alpha]$
to be the empty word.
For all $0 \leq \alpha, \beta \leq m$ we let
$\overline{[\alpha,\beta]}=[\beta,\alpha]$, then always
$\overline{w[\alpha,\beta]}=w \overline{[\alpha,\beta]}$.

Let us focus on the word $w_0 \in \Gamma_0^*$ which in our notation is the
solution $w_0 = \sigma(L_0) = \sigma(R_0)$, where
$L_0 = x_1 \cdots x_g$ and $R_0 = x_{g+1} \cdots x_d$,
$x_i \in (\Gamma_0 \cup \Omega_0)$ for $1 \leq i \leq d$. We are going to
define an equivalence relation $\approx$ on the set of intervals of $w_0$.
For this we have to fix some few more notations. We let $m_0 = |w_0|$ and
for $i \in \{1, \ldots, d\}$ we define positions
$\leftp(i) \in \{0, \ldots, m_0-1\}$ and $\rightp(i) \in \{1, \ldots, m_0\}$
by the congruences
\begin{eqnarray*}
 \leftp(i) & \equiv & |\sigma(x_1 \cdots x_{i-1})| \mod m_0, \\
 \rightp(i) & \equiv & |\sigma(x_{i+1} \cdots x_d)| \mod m_0.
\end{eqnarray*}
This means, the factor $\sigma(x_i)$ starts in $w_0$ at the left position
$\leftp(i)$ and it ends at the right position $\rightp(i)$.
In particular, we have $\leftp(1) = \leftp(g+1) = 0$ and $\rightp(g) =
\rightp(d) =m_0$. The set of $\leftp$ and $\rightp$ positions is
called the set of {\em cuts\/}. Thus, the set of cuts is $\{\,
\leftp(i), \rightp(i) \mid 1 \leq i \leq d \,\}$. There are at most
$d$ cuts. These positions cut the word $w_0$ in at most $d-1$ factors.
For convenience we henceforth assume $2 \leq g < d < m_0$ whenever
necessary. We make also the assumption that $\sigma(x_i) \neq 1$ for
all $1 \leq i \leq d$. This assumption can be realized e.g. by a first
step in Plandowski's algorithm using a partial solution $\delta$ which sends a
variable $X$ to the empty word, if $\sigma(X) =1$ and sends $X$ to
itself otherwise. Another choice to realize this assumption is by a
guess in some preprocessing.

We have $\sigma(x_i) = w_0[\leftp(i), \rightp(i)]$ and
$\sigma(\overline{x_i}) = w_0[\rightp(i), \leftp(i)]$ for $1 \leq i
\leq d$.
By our assumption, the interval $ [\leftp(i), \rightp(i)]$
is positive.
Let us consider a pair $(i, j)$ such that $i,j \in {1, \ldots, d}$ and
$x_i = x_j$ or $x_i = \overline{x_j}$. For
$\mu, \nu \in \{0, \ldots, \rightp(i)-\leftp(i)\}$ we define a relation
$\sim$ by:

\begin{eqnarray*}
 {[}\leftp(i)+\mu, \leftp(i)+\nu{]} & \sim &
 {[}\leftp(j)+\mu, \leftp(j)+\nu{]}, \mathrm{\,if\,} x_i = x_j, \\
 {[}\leftp(i)+\mu, \leftp(i)+\nu{]} & \sim &
 {[}\rightp(j)-\mu, \rightp(j)-\nu {]}, \mathrm{\,if\,} x_i = \overline{x_j}.
\end{eqnarray*}
Note that $\sim$ is a symmetric relation. Moreover,
$[\alpha, \beta] \sim [\alpha', \beta']$ implies both
$[\beta, \alpha] \sim [\beta', \alpha']$ and
$w_0[\alpha, \beta] = w_0[\alpha', \beta']$. By $\approx$ we denote the
reflexive and transitive closure of $\sim$. Then $\approx$ is an equivalence
relation and again,
$[\alpha, \beta] \approx [\alpha', \beta']$ implies both
$[\beta, \alpha] \approx [\beta', \alpha']$ and
$w_0[\alpha, \beta] = w_0[\alpha', \beta']$.

Next we define the notion of {\em free interval\/}. An interval
$[\alpha, \beta]$ is called {\em free\/}, if whenever
$[\alpha, \beta] \approx [\alpha', \beta']$, then there is no cut
$\gamma'$ with
$\min\{\alpha', \beta'\} < \gamma' < \max\{\alpha', \beta'\}$. Clearly, the
set of free intervals is closed under involution, i.e., if
$[\alpha, \beta]$ is free, then $[\beta, \alpha]$ is free, too. It is also
clear that $[\alpha, \beta]$ is free whenever $|\beta-\alpha| \leq 1$.

\begin{example}\label{ex-free}
The last equation in Example~\ref{ex-delta}, namely
\[
   a X \ox \oa = Y \ob Y \oa b \oy, 
\]
has a solution which yields  the word
\[ 
  w_0 =\;  \stackrel{0}{|} a \stackrel{1}{|}  b c \oc \ob \stackrel{5}{|}   
	\ob  \stackrel{6}{|}  a b c \stackrel{9}{|}   
	\oc \ob  \stackrel{11}{|} \oa \stackrel{12}{|} b 
	\stackrel{13}{|}  b c \oc    \ob \stackrel{17}{|} \oa \stackrel{18}{|}  .
\]
  The set of cuts is shown by the bars.
The intervals $[1,5]$, $[13,17]$, and $[6,9]$ are not free,
since $[1,5] \approx [17,13] \approx [7,11]$
 and  $[6,9] \approx [0,3]$ and $[7,11]$, $[0,3]$
contain cuts. 
There is only one equivalence class 
of free intervals   of length
longer than 1 (up to involution), which is given by 
$[1,3] \sim [17,15] \sim [7,9] \sim [11,9] \sim [5, 3] \sim [13,15]$.
\end{example}

The next lemma says that  subintervals of free intervals
are free again.

\begin{lemma} \label{freesub}
Let $[\alpha, \beta]$ be a free interval and
$\mu, \nu$ such that 
$\min\{\alpha,\beta\} \leq \mu,\nu \leq  \max\{\alpha,\beta\}$.
Then the interval $[\mu, \nu]$ is also free.
\end{lemma}

\begin{proof}
We may assume that $\alpha \leq \mu < \nu \leq \beta$. By
contradiction assume that $[\mu, \nu]$ is not free. Then there is some
$k \geq 0$ and some cut $\gamma'$ such that
$$[\mu,\nu]=[\mu_0,\nu_0] \sim [\mu_1, \nu_1] \sim \cdots \sim [\mu_k, \nu_k]$$
with $\min\{\mu_k,\nu_k\} < \gamma' < \max\{\mu_k,\nu_k\}$. If $k=0$, then
we have a immediate contradiction. For $k \geq 1$ the relation
$[\mu,\nu] \sim [\mu_1,\nu_1]$ is due to some pair $x_i$, $x_j$ with
$x_i = x_j$ or $x_i = \overline{x_j}$. Since $[\alpha, \beta]$ contains
no cut, we can use the same pair to find an interval $[\alpha_1, \beta_1]$
such that $[\alpha, \beta] \sim [\alpha_1, \beta_1]$ and
$\mu_1, \nu_1 \in \{ \min\{\alpha_1,\beta_1\}, \ldots,
\max\{\alpha_1,\beta_1\}\}$. Using induction on $k$ we see that
$[\alpha_1, \beta_1]$ cannot be free. A contradiction,
because then $[\alpha, \beta]$ is not free.\qed
\end{proof}

Next we introduce the notion of {\em implicit cut\/} for non-free
intervals. For our purpose it is enough to define it for positive intervals.
So, let $0 \leq \alpha < \beta \leq m_0$ such that $[\alpha, \beta]$ is not
free. A position $\gamma$ with $\alpha < \gamma < \beta$ is called an
{\em implicit cut\/} of $[\alpha, \beta]$, if we meet the following
situation. There is a cut $\gamma'$ and an interval $[\alpha', \beta']$ such
that
\begin{eqnarray*}
 \min\{\alpha', \beta'\} \, < & \gamma' & < \, \max\{\alpha', \beta'\}, \\
 {[}\alpha, \beta {]} & \approx & {[}\alpha', \beta'{]}, \\
 \gamma-\alpha & = & |\gamma' - \alpha'|.
\end{eqnarray*}
The following observation will be used throughout. If we have
$\alpha \leq \mu < \gamma < \nu \leq \beta$ and $\gamma$ is an implicit cut of
$[\alpha, \beta]$, then $\gamma$ is also an implicit cut of $[\mu, \nu]$.
In particular, neither $[\mu, \nu]$ nor $[\nu, \mu]$ is a free
interval.\footnote{However, 
if $\gamma$ is an implicit cut of $[\mu,\nu]$, then it might happen
that $\gamma$ is no implicit cut of $[\alpha, \beta]$, although
$[\alpha, \beta]$ is certainly not free.}

\begin{lemma} \label{mfo}
Let $0 \leq \alpha \leq \alpha' < \beta \leq \beta' \leq m_0$ such that
$[\alpha, \beta]$ and $[\alpha', \beta']$ are free intervals. Then the
interval $[\alpha, \beta']$ is free, too.
\end{lemma}

\begin{proof}
Assume by contradiction that $[\alpha, \beta']$ is not free. Then it contains
an implicit cut $\gamma$ with $\alpha < \gamma < \beta'$. By the observation
above: If $\gamma < \beta$, then $\gamma$ is an implicit cut of $[\alpha,
\beta]$ and $[\alpha, \beta]$ is not free. Otherwise, $\alpha' < \gamma$ and
$\alpha', \beta'$ is not free. \qed
\end{proof}

We now consider the maximal elements. A
free interval $[\alpha, \beta]$ is called {\em maximal free\/}, if there is
no free interval $[\alpha', \beta']$ such that both
$\alpha' \leq \min\{\alpha, \beta\} \leq \max \{\alpha, \beta\} \leq \beta'$
and $\beta'-\alpha' > |\beta-\alpha|$. With this notion Lemma~\ref{mfo}
states that maximal free intervals do not overlap.

\begin{lemma} \label{freecut}
Let $[\alpha, \beta]$ be a maximal free interval. Then there are intervals
$[\gamma, \delta]$ and $[\gamma', \delta']$ such that
$[\alpha, \beta] \approx [\gamma, \delta] \approx [\gamma', \delta']$ and
$\gamma$ and $\delta'$ are cuts.
\end{lemma}

\begin{proof}
We may assume that $[\alpha, \beta]$ is a positive interval, i.e.,
$\alpha < \beta$. We show the existence of $[\gamma, \delta]$ where
$[\alpha, \beta] \approx [\gamma, \delta]$ and $\gamma$ is a cut. The
existence of $[\gamma', \delta']$ where
$[\alpha, \beta] \approx [\gamma', \delta']$ and $\delta'$ is a cut follows
by a symmetric argument.

If $\alpha=0$, then $\alpha$ itself is a cut and we can choose
$\delta=\beta$. Hence let $1 \leq \alpha$ and consider the positive interval
$[\alpha-1,\beta]$. This interval is not free, but the only possible
position for an implicit cut is $\alpha$. Thus for some cut $\gamma$ we have
$[\alpha-1, \beta] \approx [\alpha', \beta']$ with $\min \{\alpha',
\beta'\} < \gamma < \max\{\alpha', \beta'\}$ and $|\gamma-\alpha'|=1$. A
simple reflection shows that we have
$[\alpha-1, \alpha] \approx [\alpha',\gamma]$ and
$[\alpha, \beta] \approx [\gamma, \beta']$. Hence we can choose
$\delta=\beta'$. \qed
\end{proof}

\begin{proposition} \label{nfi}
Let $\Gamma$ be the set of words $w \in \Gamma_0^*$ such that there is a
maximal free interval $[\alpha,\beta]$ with $w=w_0[\alpha,\beta]$. Then
$\Gamma$ is a subset of $\Gamma_0^+$ of size at most $2d-2$. The set
$\Gamma$ is closed under involution.
\end{proposition}

\begin{proof}
Let $[\alpha,\beta]$ be maximal free. Then $|\beta-\alpha| \geq 1$ and
$[\beta, \alpha]$ is maximal free, too. Hence $\Gamma \subseteq \Gamma_0^+$
and $\Gamma$ is closed under involution. By Lemma~\ref{freecut} we may
assume that $\alpha$ is a cut. Say $\alpha<\beta$. Then $\alpha \neq m_0$
and there is no other maximal free interval $[\alpha, \beta']$ with
$\alpha<\beta'$ because of Lemma~\ref{mfo}. Hence there are at most $d-1$
such intervals $[\alpha, \beta]$. Symmetrically, there are at most $d-1$
maximal free intervals $[\alpha, \beta]$ where $\beta<\alpha$ and $\alpha$
is a cut.\qed
\end{proof}

For a moment let $\Gamma_0'=\Gamma_0 \cup \Gamma$ where
$\Gamma \subseteq \Gamma_0^+$ is the set defined in Proposition~\ref{nfi}.
The inclusion $\Gamma_0' \subseteq \Gamma_0^+$ defines a natural
projection  $\pi: \Gamma_0' \to \Gamma_0^*$ and a mapping
$h_0': \Gamma_0' \to M$ by $h_0'=h_0 \pi$.
Consider the equation with constraints $\pi^*(E)$, this 
 is a node in the search graph, because
the size of $\Gamma$ is linear in $d$. 

The reason to switch from $\Gamma_0$ to $\Gamma_0'$ is that, due to the
constraints, the word $w_0$ may have long free intervals,  even in a
minimal solution. 
Over $\Gamma_0'$ long free intervals can be avoided. 
Formally, we replace $w_0$ by a solution $w_0'$ where
$w_0' \in \Gamma^*$. The definition of $w_0'$ is based on a factorization
of $w_0$ in maximal free intervals. There is a unique sequence
$0=\alpha_0<\alpha_1<\cdots<\alpha_k=m_0$ such that
$[\alpha_{i-1},\alpha_i]$ is a maximal free interval for 
each $1 \leq i \leq k$
and
$$w_0=w_0[\alpha_0,\alpha_i] \cdots w_0[\alpha_{k-1},\alpha_k].$$
Note
that all cuts occur as some $\alpha_p$, therefore we can think of the
factors $w_0[\alpha_{i-1},\alpha_i]$ as letters in $\Gamma$ for $1
\leq i \leq k$. Moreover, all constants which appear in $L_0R_0$ are
elements of $\Gamma$.
We replace $w_0$ by the word $w_0' \in \Gamma^*$.
Then we can define $\sigma: \Omega \to \Gamma^*$ such that both
$\sigma(L_0)=\sigma(R_0)=w_0'$ and $\rho_0 = h_0'\sigma$.  In other
terms, $\sigma$ is a solution of $\pi^*(E_0)$. We have $w_0=\pi(w_0')$
and $\pexp(w_0') \leq \pexp(w_0)$. The crucial point is that $w_0'$
has no long free intervals anymore. With respect to $w_0'$ and
$\Gamma_0'$ all maximal free intervals have length exactly one.

In the next step we  show that
we can reduce the alphabet of constants to be $\Gamma$.
The inclusion of $\Gamma$ in ${\Gamma'}_0 $
defines an
admissible base change $\beta: \Gamma \to {\Gamma'}_0 $. 
Consider 
$E_0' = (\Gamma , h , \Omega_0 , \rho_0 ; L_0=R_0)$
where $h$ is the restriction of the mapping $h_0'$.
Then we have $\pi^*(E_0) = \beta_*(E_0')$.
 The search graph contains an arc from $E_0$ to $E_0'$,
since we may choose $\delta$ to be the identity. 
The equation with constraints $E_0'$ has a solution
$\sigma$ with $\sigma(L_0)= w_0'$
and $\pexp(w_0') \leq \pexp(w_0)$.

In order to avoid too many notations we identify $E_0$ and $E_0'$,
hence we also assume $w_0=w_0'$. 
However, as a reminder that we have changed
the alphabet of constants (recall that some words became letters),
we prefer to use the notation $\Gamma $ rather than 
 $\Gamma_0$.
Thus, in what follows we shall make the following assumptions:
\begin{eqnarray*}
 E_0 & = & (\Gamma , h , \Omega_0 , \rho_0 ; L_0=R_0),\\
 L_0 & = & x_1 \cdots x_g \mbox{ and } g \geq 2, \\
 R_0 & = & x_{g+1} \cdots x_d \mbox{ and } d>g, \\
 |\Gamma|  & \leq &  2d-2, \\
|\Omega_0 | & \leq & 2d, \\
 M & \subseteq & \B^{2n \times 2n}.
\end{eqnarray*}

Moreover: All variables $X$ occur in $L_0R_0 \overline{L_0R_0}$.
There is a solution $\sigma$ such that
$w_0=\sigma(L_0)=\sigma(R_0)$ with $\sigma(X_i) \neq 1$ for
$1 \leq i \leq d$ and $\rho_0 = h\sigma =h_0\sigma$. We have $|w_0|=m_0$
and $\pexp(w_0) \in 2^{\OO(d+n\log n)}$. All maximal free intervals have length
exactly one, i.e., every positive interval $[\alpha, \beta]$ with
$\beta-\alpha>1$ contains an implicit cut.

It is because of the last sentence that we have worked out 
the details about free intervals.
This difficulty is due to the constraints. Without them
the reasoning would have been much simpler.
But the good news are that from now on, the presence of
constraints will not interfere very much.

\begin{example}\label{ex-d}
Following Example~\ref{ex-free}, we use the same equation
$ a X \ox \oa = Y \ob Y \oa b \oy$ and we consider the solution $w_0$.
 
 The new solution is defined by replacing in $w_0$ each factor
$bc$ by a new letter $d$ which represents a maximal free interval.
The new $w_0$ has the form
\[ 
  w_0 =\;  \stackrel{0}{|} a \stackrel{1}{|}  d \od \stackrel{3}{|}   
	\ob  \stackrel{4}{|}  a  d \stackrel{6}{|}   
	\od  \stackrel{7}{|} \oa \stackrel{8}{|} b 
	\stackrel{9}{|}  d \od \stackrel{11}{|} \oa \stackrel{12}{|}.
\]
Now all maximal intervals have length one.
\end{example}

\section{Critical Words and Blocks}

In the following $\ell$ denotes an integer which varies between $1$ and
$m_0$. For each $\ell$ we define the set of critical words $C_\ell$ by
$$C_\ell = \{\,w_0[\gamma-\ell,\gamma+\ell], w_0[\gamma+\ell,\gamma-\ell] \mid
\gamma {\rm\,is\,a\,cut\,and\,} \ell \leq \gamma \leq m_0 - \ell\,\}.$$
We have $1 \leq |C_\ell| \leq 2d-4$ and $C_\ell$ is closed under involution.
Each word $u \in C_\ell$ has length $2\ell$, it can be written in the form
$u=u_1 u_2$ with $|u_1|=|u_2|=\ell$.
Then $u_1$ (resp. $\overline{u_2}$) appears as a suffix,
 left of some  cut and $u_2$ (resp. $\overline{u_1}$) appears as a prefix,
 right of the same  cut.

A triple $(u,w,v) \in (\{1\} \cup \Gamma^\ell) \times \Gamma^+ \times
(\{1\} \cup \Gamma^\ell)$
is called a {\em block\/} if first, up to a possible prefix or suffix
no other factor of the word $uwv$ is a critical word, second, $u \neq
1$
if and only if
a prefix of $uwv$ of length $2\ell$ belongs to $C_\ell$, and third,
$v \neq 1$ if and only if a suffix of $uwv$ of length $2\ell$ belongs to $C_\ell$.
The set of blocks is denoted by $B_\ell$. It is viewed (as a possibly
infinite) alphabet where the involution is defined by
$\overline{(u,w,v)}=(\overline{v}, \overline{w}, \overline{u})$. We can
define a homomorphism $\pi_\ell: B_\ell^* \to \Gamma^*$ by
$\pi_\ell(u,w,v)=w \in \Gamma^+$ being extended
to a projection  $\pi_\ell: (B_\ell \cup \Gamma)^* \to
\Gamma^*$
by leaving $\Gamma$ invariant. We define $h_\ell:(B_\ell \cup \Gamma) \to M$
by $h_\ell = h  \pi_\ell$. In the following we shall consider finite subsets
$\Gamma_\ell \subseteq B_\ell \cup \Gamma$ which are closed under involution. Then by
$\pi_\ell: \Gamma_\ell^* \to \Gamma^*$ and $h_\ell: \Gamma_\ell^* \to M$
we understand the restrictions of the respective homomorphisms.

For every non-empty word $w \in \Gamma^+$ we define its
{\em $\ell$-factorization\/} as follows. We write
$$F_\ell(w) = \Fl 1k \in B_\ell^+$$
such that $w=w_1 \cdots w_k$ and for $1 \leq i \leq k$ the following
conditions are satisfied:
\begin{itemize}
\item $v_i$ is a prefix of $w_{i+1} \cdots w_k$,
\item $v_i=1$ if and only if $i=k$,
\item $u_i$ is a suffix of $w_1 \cdots w_{i-1}$,
\item $u_i=1$ if and only if $i=1$.
\end{itemize}
Note that the $\ell$-factorization of a word $w$ is unique. For
$k\geq2$ we 
have $|w_1|\geq \ell$ and $|w_k| \geq \ell$, but all other $w_i$ may
be short. If no critical word appears as a factor of $w$, then
$F_\ell(w) = (1,w,1)$. In particular, this is the case for
$|w|<2\ell$.  If we have $w=puvq$ with $|u|=|v|=\ell$ and $uv \in
C_\ell$, then there is a unique $i \in \{1, \ldots, k-1\}$ such that
$u=u_{i+1}$, $v=v_i$, and $pu=w_1 \cdots w_i$, $vq= w_{i+1} \cdots
w_k$. Thus, $F_\ell(w)$ contains a factor $(u_i, w_i, v) (u, w_{i+1},
v_{i+1})$ where $v$ is a prefix of $w_{i+1}v_{i+1}$ and $u$ is a
suffix of $u_iw_i$. For example, the $\ell$-factorization of $uv \in
C_\ell$ with $|u|=|v|=\ell$ is
$$F_\ell(uv)=(1,u,v)(u,v,1).$$

We define the $\head$, $\body$, and $\tail$ of a word $w$ based on its
$\ell$-factorization
$$F_\ell(w)= \Fl 1k$$
in $B_\ell^*$ and $\Gamma^*$ as follows:
\begin{eqnarray*}
 \Head_\ell(w) & = & \uwv 1 \in B_\ell, \\
 \head_\ell(w) & = & w_1 \in \Gamma^+, \\
 \Body_\ell(w) & = & \Fl 2{k-1} \in B_\ell^*, \\
 \body_\ell(w) & = & w_2 \cdots w_{k-1} \in \Gamma^*, \\
 \Tail_\ell(w) & = & \uwv k \in B_\ell, \\
 \tail_\ell(w) & = & w_k \in \Gamma^+.
\end{eqnarray*}
For $k \geq 2$ (in particular, if $\body_\ell(w) \neq 1$) we have
\begin{eqnarray*}
F_\ell(w)  & = & \Head_\ell(w) \Body_\ell(w) \Tail_\ell(w),\\
w & = &  \head_\ell(w) \body_\ell(w) \tail_\ell(w).
\end{eqnarray*}
 Moreover, $u_2$ is a
suffix of $w_1$ and $v_{k-1}$ is a prefix of $w_k$.

Assume $\body_\ell(w) \neq 1$ and let $u,v\in \Gamma^*$ be any words.
Then we can view $w$ in the context $uwv$ and $\Body_\ell(w)$ appears
as a proper factor in the $\ell$-factorization of $uwv$. More
precisely, let
$$F_\ell(uwv) = \Fl 1k.$$
Then there are unique $1 \leq p <q \leq k$ such that:
\begin{eqnarray*}
 F_\ell(uwv) & = & \Fl 1p \Body_\ell(w) \Fl qk,\\
 w_1 \cdots w_p & = & u \, \head_\ell(w), \\
 w_q \cdots w_k & = & \tail_\ell(w) v
\end{eqnarray*}
Finally, we note that the above definitions are compatible with the
involution. We have $ F_\ell(\overline{w}) = \overline{F_\ell(w)}$, $
\Head_\ell(\overline{w}) = \overline{\Tail_\ell(w)}$, and
$\Body_\ell(\overline{w}) = \overline{\Body_\ell(w)}$.

\section{The $\ell$-Transformation}

Our equation with constraints is
$E_0 = (\Gamma, h, \Omega_0, \rho_0; x_1 \cdots x_g = x_{g+1} \cdots x_d).$
We start with
the $\ell$-factorization of $w_0=\sigma(x_1 \cdots x_g) =
\sigma(x_{g+1} \cdots x_d)$. Let
$$F_\ell(w_0) = \Fl 1k.$$
A sequence $S = \Fl pq$ with $1 \leq p \leq
q \leq k$ is called an {\em $\ell$-factor\/}. We say that $S$ is a
\emph{cover} of a positive interval $[\alpha,\beta]$, if both $|w_1
\cdots w_{p-1}| \leq \alpha$ and $|w_{q+1} \cdots w_k|\leq m_0-\beta$.
Thus, $w_0[\alpha,\beta]$ becomes a factor of $w_p \cdots w_q$. 
It is
 a {\em minimal cover\/}, if neither $\Fl {p+1}{q}$ nor $\Fl
p{q-1}$ is a cover of $[\alpha,\beta]$. The minimal cover exists and
it is unique.

We let $\Omega_\ell = \{\,X \in \Omega_0 \mid
\body_\ell(\sigma(X))\neq 1\,\}$, and we are going to define a new
left-hand side $L_\ell \in (B_\ell \cup \Omega_\ell)^*$ and a new
right-hand side $R_\ell \in (B_\ell \cup \Omega_\ell)^*$. For $L_\ell$
we consider those $1 \leq i \leq g$ where $\body_\ell(\sigma(x_i))
\neq 1$. Note that this implies $x_i \in \Omega_\ell$ since $\ell \geq
1$ and then the $\body$ of a constant is always empty. Recall the
definition of $\leftp(i)$ and $\rightp(i)$, and define
$\alpha=\leftp(i)+|\head_\ell(\sigma(x_i))|$ and $\beta = \rightp(i) -
|\tail_\ell(\sigma(x_i))|$. Then we have $w_0[\alpha,\beta]=
\body_\ell(\sigma(x_i))$. Next consider the $\ell$-factor $S_i = \Fl
pq$ which is the minimal cover of $[\alpha,\beta]$. Then we have $1<p
\leq q < k$ and $w_p \cdots w_q = w_0[\alpha,\beta] =
\body_\ell(\sigma(x_i))$.
The definition of $S_i$ depends only
on $x_i$, but not on the choice of the index $i$.

We replace the
$\ell$-factor $S_i$ in $F_\ell(w_0)$ by the variable $x_i$.  Having
done this for all $1 \leq i \leq g$ with $\body_\ell(\sigma(x_i)) \neq
1$ we obtain the left-hand side $L_\ell \in (B_\ell \cup
\Omega_\ell)^*$ of the $\ell$-transformation $E_\ell$. For $R_\ell$ we
proceed analogously by replacing those $\ell$-factors $S_i$ where
$\body_\ell(\sigma(x_i)) \neq 1$ and $g+1 \leq i \leq d$.

For $E_\ell$ we cannot use the alphabet $B_\ell$, because it might be too large or
even infinite. Therefore we let $\Gamma_{\ell'}$ be the smallest subset
of $B_\ell$ which is closed under involution and which satisfies
$L_\ell R_\ell \in (\Gamma_{\ell'} \cup \Omega_\ell)^*$. We let
$\Gamma_\ell = \Gamma_{\ell'} \cup \Gamma$. The projection $\pi_\ell :
\Gamma_\ell^* \to \Gamma^*$ and the mapping $h_\ell : \Gamma_\ell \to M$ are
defined by the restriction of $\pi_\ell: B_\ell \to \Gamma^*$,
$\pi_\ell(u,w,v)=w$ and $h_\ell(u,w,v)=h(w) \in M$ and by
$\pi_\ell(a)=a$ and  $h_\ell(a)=h(a)$ for $a \in \Gamma$.

Finally, we define the mapping $\rho_\ell : \Omega_\ell \to M$ by
$\rho_\ell(X) = h(\body_\ell(\sigma(X)))$. This completes the
definition of the $\ell$-transformation:
$$E_\ell = (\Gamma_\ell , h_\ell, \Omega_\ell, \rho_\ell; L_\ell = R_\ell).$$

\begin{remark}
One can  verify that
$\sigma_\ell: \Omega_\ell \to \Gamma_\ell^*$,
$\sigma_\ell(X) = \phi_\ell(\Body_\ell(\sigma(X)))$ defines a solution of
$E_\ell$, where $\phi_\ell$ is the identity on $\Gamma_\ell$ and $\pi_\ell$
on $B_\ell \setminus \Gamma_{\ell'}$. Although, up to the trivial case $\ell=m_0$,
we make no explicit use of this fact.
\end{remark}

\begin{example}
We continue with our example
$a X \ox \oa = Y \ob Y \oa b \oy$ and the solution $\sigma$ which has
been given by
\[
  w_0 =\;  \mid a \mid  d \od \mid   
	\ob  \mid  a  d \mid   
	\od  \mid \oa \mid b 
	\mid  d \od \mid \oa \mid,
\]
where the bars show the cuts.

Up to involution,
  the set $C_{1}$ is given by $\{ ad, bd, \oa b, d \od \}$ and
$C_2$ is given by  $\{ d\od \ob a, \od \ob a d, a d \od \oa, d \od \oa b \}$.
The 1-factorization of $w_0$ can be obtained letter by letter.
 The 2-factorization of $w_0$ is given by the following sequence:
\[
   (1,a d \od, \ob a) (d \od, \ob, ad) (\od \ob, a d, \od \oa)
	  (a d, \od, \oa b) ( d \od, \oa, bd) (\od \oa, b, d \od)
	(\oa b, d \od a,1).
\]

Recall $\sigma(X)= d \od \ob a d $ and $\sigma(Y)=a d \od$. 
Hence their 2-factorizations are
 $ (1,a d \od, \ob a) (d \od, \ob, ad) (\od \ob, a d, 1)$ and
$(1, a d \od, 1)$, respectively.

 By renaming letters, the 2-factorization of $w_0$ becomes
 $a \ob c d e b \oa$ and the equation $E$ reduces to
$E_2: aXcdeX\oa = a \ob c d e b \oa$ since the body
of $\sigma(Y)$ is empty.

  The reader can check that the 3-factorization of $w_0$ after 
renaming is the very same word as the 2-factorization, but the
3-factorization of $\sigma(X)$ is now one letter, 
$(1, d \od \ob a d, 1)$,
 so $E_3$
becomes a trivial equation. Plandowski's algorithm will return
{\em true} at this stage.
\end{example}

\begin{remark} \label{extremal}
i) In the extreme case $\ell=m_0$, the $\ell$-transformation becomes
trivial. Let $a=(1,w_0,1)$. Then $\overline{a}=(1,\overline{w_0},1)$ and
$\Gamma_{m_0} = \{a,\overline{a}\} \cup \Gamma$. Moreover, we have
$L_{m_0} = R_{m_0} = a$, and $h_{m_0}(a) = h(w_0) \in M$. Since
$\Omega_{m_0}= \emptyset$, the equation with constraints $E_{m_0}$ has
trivially a solution. It is clear that $E_{m_0}$ is a node in the search
graph, and if we reach $E_{m_0}$, then the algorithm will return
{\em true\/}.

ii) The other extreme case is $\ell=1$. The situation again is simple,
but the precise definition is technically more involved.
Consider a block $(u,w,v)$ which appears in $F_1(w_0)$. Then
$w=w_0[\alpha, \beta]$ for some $\beta-\alpha \geq 1$. We cannot have
$\beta - \alpha \geq 2$, because then $[\alpha, \beta]$ would have
an implicit cut $\gamma$, but $w_0[\gamma-1, \gamma+1] \in C_1$
and no critical word is a factor of $w$. An immediate
consequence is $|\Gamma_1| \leq (|\Gamma|+1)^3 \in \OO(d^3)$. Let
$X \in \Omega_0$. Then $\Body_1(\sigma(X)) \neq 1$ if and only if
$|\sigma(X)| \geq 3$. Thus, for $X \in \Omega_1$ we have
$\sigma(X) = bcu =vde$ with $b,c,d,e \in \Gamma$ and $u,v \in \Gamma^+$.
It follows:
$$F_1(\sigma(X)) = (1,b,c)(b,c,v_2)\cdots(u_{|v|+1},d,e)(d,e,1).$$
For example, for  $|v|=1$ this means $b=u_{|v|+1}$, $c=d$, and $v_2=e$.

 We can describe
$L_1 \in \Gamma_1^*$ as follows:

For $1 \leq i \leq g$ let $w_i=\sigma(x_i)$
and $a_i$ the last letter of $\sigma(x_{i-1})$ if $i>1$ and $a_1=1$.
Let $f_i$ the first letter of $\sigma(x_{i+1})$ if $i<g$ and $f_g=1$. Let
$b_i$ the first letter of $w_i$ and $e_i$ the last letter of $w_i$.

For $|w_i|=1$ we replace $x_i$ by the $1$-factor $(a_i,b_i,f_i)$.

For $|w_i|=2$ we replace $x_i$ by the $1$-factor $(a_i,b_i,e_i)(b_i,e_i,f_i)$.

For $|w_i|\geq3$ we let $c_i$ be the second letter of $w_i$ and $d_i$ its
second last. In this case we replace $x_i$ by $(a_i,b_i,c_i)x_i(d_i,e_i,f_i)$.

The definition of $R_1$ is analogous. Thus, we obtain
$|L_1R_1| \leq 3|L_0 R_0| = 3d$, and $E_1$ is admissible. We also see that
there was an overestimation of the size of $|\Gamma_1|$. For each $x_i$ we
need at most two constants together with their involutions. Since
$\Gamma_1$ contains also $\Gamma$, we obtain
$|\Gamma_1|\leq 6d$.
\end{remark}

By the remark above, $E_1$ and $E_{m_0}$ are admissible and hence
nodes of the search graph. The goal is to reach $E_{m_0}$
via $E_1$ when starting with $E_0$.  For the
moment it is even not clear that the $\ell$-transformations with $1 <
\ell < m_0$ belongs to the search graph. We prove this statement in
the next section.

\section{The $\ell$-transformation $E_\ell$ is admissible}

\begin{proposition} \label{ladm}
There is a  polynomial $p$ (of degree at most 4) such that each $E_\ell$
is admissible for all $\ell \geq 1$.
\end{proposition}

\begin{proof}
It is enough to show that $L_\ell$ and $R_\ell$ can be represented by
exponential expressions of size $\OO(d^2(d+n\log n))$. Then $\Gamma_\ell$ can have
size at most $\OO(d^2(d+n\log n))$ and the assertion follows. We will estimate the
size of an exponential expression for $L_\ell$, only.

We start again with the $\ell$-transformation of
$$F_\ell(w_0) = \Fl 1k.$$
If $k$ is small there is nothing to do since $|L_\ell| \leq |F_\ell(w_0)|$.
An easy reflection shows that $|L_\ell|$ can become large, only if there is
some $1 \leq i \leq g$ such that $\head_\ell(\sigma(x_i))$ or
$\tail_\ell(\sigma(x_i))$ is long. By symmetry we treat the case
$\head_\ell(\sigma(x_i))$ only and we fix some notation. We let
$1 \leq i \leq g$, $\alpha = \leftp(i)$, and
$\beta = \alpha+|\head_\ell(\sigma(x_i))|$. Let
$$\Fl{p-1}{q+1}$$
be a minimal cover of $[\alpha,\beta]$. We may assume that $q-p$ is large.
It is enough to find an exponential expression for the
$\ell$-factor
$$\Fl{p}{q}$$
having size in $\OO(d(d+n\log n))$, because we want the whole expression
to have size in $\OO(d^2(d+n\log n))$.

Note that $w_p \cdots w_q$ is a proper factor of $\head_\ell(\sigma(x_i))$.
Hence no critical word of $C_\ell$ can appear as a factor inside
$w_p \cdots w_q$. This means there is some $p \leq s \leq q$ such that
both $|w_p \cdots w_{s-1}| < \ell$ and $|w_{s+1} \cdots w_q| < \ell$. Indeed,
if $|w_p \cdots w_{q-1}| < \ell$, then we choose $s=q$. Otherwise we let
$p\leq s \leq q$ be minimal such that $|w_p \cdots w_s| \geq \ell$. Then
$|w_{s+1} \cdots w_q| \geq \ell$ is impossible because
$u_{s+1}v_s \in C_\ell$ would appear as a factor in $w_p \cdots w_q$. We can
write
$$ \Fl pq = S_1 \uwv s S_2;$$
and since $\uwv s \in \Gamma_\ell$ is a letter, it is enough to find
exponential expressions for $S_i$, $i=1,2$, of size $\OO(d(d+n\log n))$ each.
As a conclusion it is enough to prove the following lemma. \qed
\end{proof}

The statement of the next lemma is slightly more general as we need it
above. There we need the lemma for  $c = 1$, but later we
will apply the lemma with values $c \leq 32 d$.

\begin{lemma} \label{short}
Let $c > 0$ be a number and
$$S=\Fl{1}{k} \in B_\ell^*$$
be a sequence which appears as some $\ell$-factor in $F_\ell(w_0)$. If we
have $k \leq 3$ or $|w_2 \cdots w_{k-1}| \leq c\ell$, then we can represent
the sequence by some exponential expression of size $\OO(cd(d+n\log n))$.
\end{lemma}

\begin{proof}
  We show that there is an exponential expression of size
  $\OO(d(d+n\log n))$ under the assumption $|w_1 \cdots w_k|<\ell$.
  This is enough, because we always can write $S$ as $a_0S_1 a_1
  \cdots S_{c'} a_{c'}$, where $c'\leq c $, the $a_i$ are letters, and
  each $S_i$ satisfies the assumption.  Note that the assumption
  implies $u_1 \neq 1 \neq v_k$
and we may define $u_{k+1}$ as the suffix
of length $\ell$ of $u_1w_1 \cdots w_k$. For $1 \leq i \leq k$ let
  $z_i=u_{i+1}v_i$. Then $z_i \in C_\ell$ is a critical word which
  appears as a factor in $z = u_1 w_1 w_2 \cdots w_k v_k$. If the
  words $z_i$, $1 \leq i <k$ are pairwise different, then $k-1 \leq
  |C_\ell| \in \OO(d)$ and we are done.  Hence we may assume that
  there are repetitions. Let $j$ be the smallest index such that a
  critical word is seen for the second time and let $i<j$ be the first
  appearance of $z_j$.  This means for $1 \leq i < j$ the words $z_1,
  \cdots, z_{j-1}$ are pairwise different and $z_i=z_j$. Now, $|w_1
  \cdots w_k|<\ell$ and $|z_i|=2\ell$, hence $z_i$ and $z_j$ overlap
  in $z$.  We can choose $r$ maximal such that $u_1 w_1 \cdots w_i
  (w_{i+1} \cdots w_j)^rv_j$ is a prefix of the word $z$.  (Note that
  the last factor $v_j$ insures that the prefix ends with $z_j$).  For
  some index $s>j$ we can write
$$
z=u_1 w_1 \cdots w_i (w_{i+1} \cdots w_j)^r w_s \cdots w_k v_k.$$
We claim that $z_i \not \in \{z_s, \ldots, z_k\}$.  Indeed, 
let $t$ be maximal such  that $z_i = z_t$ and assume that $j\not= t$. Then both
$|w_{i+1} \cdots w_{j}|$ and $|w_{j+1} \cdots w_{t}|$ are periods of
$z_i$, but $|w_{i+1} \cdots w_{t}| \leq |z|$.  Hence by Fine and
Wilf's Theorem \cite{Lothaire83} we obtain that the greatest common divisor
of $|w_{i+1} \cdots w_{j}|$ and $|w_{j+1} \cdots w_{t}|$
is a period, too. Due to the definition of an $\ell$-factorization 
($z_j$ was the first repetition)   the length $|w_{j+1} \cdots w_{t}|$
is therefore a multiple of $|w_{i+1} \cdots w_{j}|$ and
we must have $t = s-1$. This shows the claim.  Moreover, we have
\begin{eqnarray*}
 & & \Fl 1k \\
 & = & \Fl 1i [\Fl {i+1} j]^r\, S'
\end{eqnarray*}
where $S' =  \Fl sk$ for $ s = i +1 + r(j-i)$.
We have $r \leq \pexp(w_0)$, hence $r \in 2^{\OO(d+n\log n)}$. It follows that
$$\Fl 1i[\Fl {i+1}j]^r$$
is   an  exponential expression of  size  $j+
\log(r) \in   \OO(d+n\log   n)$. More precisely,    for  some suitable
constant $\widetilde c$ its size is at most $\widetilde c(d+n\log n)$.
The constant $\widetilde   c$ depends only on the constant which  is
hidden when  writing $\exp(w_0)  \in  2^{\OO(d+n\log
  n)}$.  By induction on the size of the set $\{z_1, \ldots, z_k\}$ we
may assume that $S'= \Fl sk$ has  an exponential expression of size at
most  $|\{z_s,    \ldots, z_k\}|    \widetilde c   (d+n)$.   Hence the
exponential expression for $S$ has size at most
$$\widetilde c (d+n\log n ) + |\{z_s, \ldots, z_k\}|\widetilde c
(d+n\log n) \leq |\{z_1, \ldots, z_k\}|\widetilde c (d+n\log n).$$
Thus, the size is in $\OO(d(d+n\log n))$. \qed
\end{proof}

At this stage we know that all $\ell$-transformations are admissible
(with respect to some suitable polynomial of degree 4). Thus $E_1, \ldots, E_{m_0}$
are nodes of the search graph. Next we show that the search graph contains
arcs $E_0 \rightarrow E_1$ and $E_\ell \rightarrow E_{{\ell'}}$ for
$1 \leq \ell < {\ell'} \leq 2\ell$. Hence the graph contains a path
(of logarithmic length in $m_0$) from $E_0$ to $E_{m_0}$. The
non-deterministic procedure is able to find this path and on input
$E_0$ Plandowski's algorithm gives the correct answer.

In order to establish the existence of arcs from $E_\ell$ to
$E_{{\ell'}}$ for $0 \leq \ell < {\ell'} \leq \max \{1, 2\ell\}$ we shall
define intermediate equations $E_{\ell,{\ell'}}$ such that there is an
admissible base change $\beta$, a projection $\pi$, and a partial solution $\delta$
with
$$\delta_*(\pi^*(E_\ell)) \equiv E_{\ell,{\ell'}} \equiv \beta_*(E_{{\ell'}}).$$

\section{The arc from $E_0$ to $E_1$}

Recall the definition of
$E_1=(\Gamma_1 , h_1, \Omega_1, \rho_1; L_1 = R_1).$
The letters of $\Gamma_1$ can be
written either as $(a,b,c)$ or as $b$ with $a,c \in \Gamma \cup \{1\}$
and $b\in \Gamma$. We define a projection which is used here as a base
change $\beta: \Gamma_1 \to \Gamma$ by $\beta(a,b,c)=b$ and leaving
the letters of $\Gamma$ invariant. Clearly, $h_1=h\beta$, and $\beta$
defines an admissible base change. Define $E_{0,1}=\beta_*(E_1)$. Then
we have $L_{0,1}=\beta(L_1)$ and $R_{0,1}=\beta({R_1})$ where $\beta:
(\Gamma_1 \cup \Omega_1)^* \to (\Gamma \cup \Omega_1)^*$ is the extension
with $\beta(X)=X$ for all $X \in \Omega_1$.  We have $\Gamma_{0,1} =
\Gamma$

It is now obvious how to define the partial solution
$\delta: \Omega_0 \to \Gamma \Omega_1 \Gamma \cup \Gamma^*$
such that $\delta_*(E_0) = E_{0,1}$. If $|\sigma(X)| \leq 2$, then we let
$\delta(X) = \sigma(X)$. For $|\sigma(X)|\geq 3 $ we write $\sigma(X)=aub$
with $a,b \in \Gamma$ and $u\in \Gamma^+$. Then we have
$X \in \Omega_1 = \Omega_{0,1}$ and we define $\delta(X)=aXb$ and
$\rho_{0,1}(X)=h(u)$. For $X \in \Omega_1$ we have
$\rho_1(X)=h(\body_1(\sigma(X)))$, hence $\rho_{0,1}=\rho_1$, too.
This shows that, indeed, $\delta_*(E_0)=\beta_*(E_1)$.
Formally, we can write this as $\delta_*(\pi^*(E_0)) =\beta_*(E_1)$,
where $\pi$ is the identity. Hence there is an arc from $E_0$ to
$E_1$.

\section{The equations $E_{\ell,{\ell'}}$ for $1\leq\ell<{\ell'}\leq 2\ell$}

In this section we define for each  $1\leq\ell<{\ell'}\leq 2\ell$ 
an intermediate equation with constraints
$$ \beta_*(E_{\ell'}) = 
E_{\ell,{\ell'}}=(\Gamma_{\ell,{\ell'}}, h_{\ell,{\ell'}}, \Omega_{{\ell'}}, \rho_{{\ell'}} ; L_{\ell,{\ell'}} = R_{\ell,{\ell'}})$$
by some base change $\beta : \Gamma_{{\ell'}} \to (B_\ell \cup \Gamma)^*$, then we show
that $\beta$ is admissible.

 Recall $\Gamma \subseteq \Gamma_{\ell'} \subseteq B_{\ell'} \cup \Gamma$.
The base change $\beta$ leaves
the letters of $\Gamma$ invariant.
Consider some $(u,w,v) \in \Gamma_{{\ell'}}\setminus \Gamma$. It is enough to define
$\beta(u,w,v)$ or $\beta(\overline{v},\overline{w},\overline{u})$.
Hence we may assume that $(u,w,v)$ appears as a letter in the
${\ell'}$-factorization $F_{{\ell'}}(w_0)$. Therefore we find a positive
interval $[\alpha,\beta]$ such that $w=w_0[\alpha,\beta]$ and such
that the following two conditions are satisfied:

1) We have $u=1$ and $\alpha=0$ or
$|u| = {\ell'}$, $\alpha \geq {\ell'}$, and $u = w_0[\alpha-{\ell'},\alpha]$.

2) We have $v=1$ and $\beta=m_0$ or $|v| = {\ell'}$, $\beta \leq m_0-{\ell'}$,
and $v = w_0[\beta, \beta+{\ell'}]$.

Let $\Fl pq$ be the $\ell$-factor which is the  minimal cover of $[\alpha,\beta]$
with respect to the $\ell$-factorization $F_\ell(w_0)$. Since
$\ell \leq {\ell'}$ we have $w_p \cdots w_q= w$. Moreover, the word $u_p$
is a suffix of $u$ and $v_q$ is a prefix of $v$. We define
$$\beta(u,w,v) = \Fl pq \in B_\ell^+.$$
The definition does not depend on the choice of $[\alpha,\beta]$ as long
as $0 \leq \alpha < \beta \leq m_0$ and 1) and 2) are satisfied. We have
$\overline{\beta(u,w,v)} = \beta(\overline{v},\overline{w},\overline{u})$
and $h_\ell\beta = h_{{\ell'}}$. Now let
$\Gamma_{\ell,{\ell'}} \subseteq B_\ell \cup \Gamma$ be the smallest subset such that
$\beta(\Gamma_{{\ell'}}) \subseteq \Gamma_{\ell,{\ell'}}^*$. Then
$\Gamma_{\ell, {\ell'}}$ contains $\Gamma$ and it is closed under involution (since $\Gamma_{{\ell'}}$
has this property). A crucial, but easy reflection shows that
$\Gamma_\ell \subseteq \Gamma_{\ell, {\ell'}}$.
This will become essential later.

We view $\beta$ as a homomorphism $\beta: \Gamma_{\ell'}^* \to
\Gamma_{\ell, {\ell'}}^*$ and define $E_{\ell, {\ell'}} =
\beta_*(E_{{\ell'}})$.  Let us show that $\beta$ defines an admissible
base change.  Since $E_{{\ell'}}$ is already known to be admissible with
respect to some  polynomial of degree 4, it is enough to find some admissible
exponential expression (again with respect to some polynomial of degree 4)
for the $\ell$-factor
$$\beta(u,w,v)= \Fl pq$$ where $(u,w,v) \in \Gamma_{{\ell'}}\setminus
\Gamma$.
 We use the same notations as above. Thus, for
some positive interval $[\alpha,\beta]$ we have
$w_p \cdots w_q=w_0[\alpha,\beta]$, the word $u$ is a suffix of
$w_0[0,\alpha]$, and $v$ is a prefix of $w_0[\beta,m_0]$. If $q-p$ is small,
there is nothing to do. By Lemma~\ref{short} we may also assume that
$\beta-\alpha > 32 d \ell$. We are to define inductively a sequence of
positions
$$\alpha=\alpha_0<\alpha_1<\cdots<\alpha_i<\cdots<\beta_i<\cdots<\beta_1<\beta_0=\beta.$$
Each time we let $W_i=w_0[\alpha_i,\beta_i]$. Thus, $W_0=w_p \cdots w_q$.
Assume that $W_i=w_0[\alpha_i,\beta_i]$ is already defined
such that  $\beta_i-\alpha_i \geq 2$.
The interval $[\alpha_i,\beta_i]$ is not free. Hence, there is some
implicit cut $\gamma_i$ with $\alpha_i < \gamma_i <\beta_i$. The word
$W_i$ is a factor of $w$, hence no factor of $W_i$ belongs to the set of
critical words $C_{{\ell'}}$. This implies $\beta_i-\gamma_i<{\ell'}$ or
$\gamma_i-\alpha_i<{\ell'}$. If we have $\beta_i-\gamma_i<{\ell'}$ then we let
$\alpha_{i+1}=\alpha_i$ and $\beta_{i+1}=\gamma_i$. In the other case we
let $\alpha_{i+1}=\gamma_i$ and $\beta_{i+1}=\beta_i$. Thus $W_{i+1}$ is
defined such that $W_{i+1}$ is a proper factor of $W_i$ with
$|W_i| - |W_{i+1}| < {\ell'}$.

We need some additional book keeping. We define $r_i \in \{\leftp,\rightp\}$
by $r_i=\rightp$ if $\beta_i=\beta_{i+1}$ and $r_i=\leftp$ otherwise
(i.e., $\alpha_i=\alpha_{i+1}$). Furthermore the implicit cut $\gamma_i$
corresponds to some real cut $\gamma_i'$ and $\alpha_i'<\gamma_i'<\beta_i'$
such that $W_i=w_0[\alpha_i',\beta_i']$ or $W_i=w_0[\beta_i',\alpha_i']$.
We define $s_i \in \{+,-\}$ by $s_i=+$ if $W_i=w_0[\alpha_i',\beta_i']$ and
$s_i=-$ otherwise (in particular, $s_i=-$ implies
$\overline{W_i} = w_0[\alpha_i',\beta_i']$). The triple $(\gamma_i',r_i,s_i)$
is denoted by $\gamma(i)$. There are at most $4(d-2)$ such triples and
$\gamma(i)$ is defined whenever $W_{i+1}$ is defined. We stop the induction
procedure after the first repetition. Thus we find $0 \leq i < j < 4d$
such that $\gamma(i)=\gamma(j)$. We obtain a sequence
$W_0, W_1, \ldots, W_i, \ldots, W_j$ where each word is a proper factor of the
preceding one. We have $|W_0|-|W_j|<4d{\ell'} \leq 8d\ell$ and due to
$|W_0|>32 d\ell$ the sequence above really exists, moreover $|W_j| >8d\ell$.

Next, we show that $W_j$ has a non-trivial overlap with itself. We treat the
case $\gamma(i)=\gamma(j)=(\gamma,\rightp,+)$ only.
The other three cases $(\gamma,\rightp,-)$, $(\gamma,\leftp,+)$,
and $(\gamma,\leftp,-)$ can be treated analogously.
For some $\alpha' < \gamma <\beta'$ we have $W_i=w_0[\alpha',\beta']$
and $W_{i+1}=w_0[\gamma,\beta']$. Thus, for some
$\gamma \leq \mu < \nu \leq \beta'$ we have $W_j=w_0[\mu,\nu]$ and we can
assume that $\mu-\gamma<(j-i){\ell'}\leq 4d{\ell'}-{\ell'} \leq 8d\ell-{\ell'}$.
On the other hand we have $\gamma(j)=(\gamma,\rightp,+)$, too. Hence
for some $\mu'<\gamma<\nu'$ with $\gamma-\mu'<{\ell'}$ we have
$W_j = w_0[\mu',\nu']$, too. Therefore $0 <\mu-\mu'<8d\ell$ and 
$W_j$ has some non-trivial overlap. We can write $W_j=W^e W'$ such that
$1 \leq |W| <8d\ell$ and $W'$ is a prefix of $W$.

Putting everything together, we arrive in all cases at a factorization
$W_0=UW^eV$ with $e \leq \pexp(w_0)$, $1 \leq |W| <8d\ell$, and
$|U|+|V|<16d\ell$. However, we have not finished yet. Recall that we are 
looking for an admissible exponential expression for
$$\beta(u,w,v) = \Fl pq.$$
Due to $|W_0|>\ell$ we can choose $r$ minimal, $p<r \leq q+1$, and $s$
maximal $p-1 \leq s < q$ such that $|w_p \cdots w_{r-1}|>|U|+\ell$
and $|w_{s+1} \cdots w_q| > |V|+\ell$. By Lemma~\ref{short} we may assume
$r<s$ and it is enough to find an exponential expression for
$$S=\Fl rs.$$
Note that the word $u_r w_r w_{r+1} \cdots w_s v_s$ is a factor of
$W^e$.
Again, we may assume that $w_r w_{r+1} \cdots w_s > 32 d\ell$.
By switching to some conjugated word $W'$ if necessary, we may assume that
$u_r w_r w_{r+1}\cdots w_s v_s$ is a prefix of $W^e$. Moreover, by
symmetry we may choose a positive interval $[\alpha,\beta]$ such that
$w_0[\alpha,\beta]=u_r w_r w_{r+1} \cdots w_s v_s$. Clearly, we
have $w_0[i,j] = w_0[i+|W|,j+|W|]$ for all $\alpha \leq i < j \leq \beta-|W|$.
In particular, the critical word $w_0[\alpha, \alpha+2\ell]$ appears as
$w_0[\alpha+|W|, \alpha+|W|+2\ell]$ again. This means that there is some
$r \leq t < s$ such that $|w_r \cdots w_t|=|W|$. More precisely, we can
choose $r\leq t < t' \leq s$ and a maximal $e'\leq e$ such that
  $$S = \bigl( \Fl rt \bigr)^{e'}\Fl {t'}s.$$
Since it holds $e'\leq \pexp(w_0)$, $|w_r \cdots w_t|=|W|$, and
$|w_{t'}\cdots w_s|\leq |W|$, the existence of an admissible
exponential expression for $\beta(u,w,v)$
follows. Hence $\beta$ is an admissible
base change.

\section{Passing from $E_\ell$ to $E_{\ell,{\ell'}}$ for $1\leq \ell <
  {\ell'} \leq 2\ell$}

In the final step we have to show that there exists some projection $\pi:
\Gamma_{\ell,{\ell'}}^* \to \Gamma_\ell^*$ and some partial solution
$\delta:\Omega_{\ell} \to \Gamma_{\ell,{\ell'}}^* \Omega_{{\ell'}}
\Gamma^*_{\ell,{\ell'}} \cup \Gamma^*_{\ell,{\ell'}}$ such that
$\delta_*(\pi^*(E_\ell)) \equiv E_{\ell,{\ell'}}$. We don't
have to care about admissibility anymore.

For the projection we have to consider a letter in
$\Gamma_{\ell,{\ell'}} \setminus \Gamma_\ell$.  Such a letter has the
form $(u,w,v)\in B_\ell$ and we may define $\pi(u,w,v) = w$ since
$\Gamma \subseteq \Gamma_\ell$.

  Clearly
$\pi(\overline{(u,w,v)}) = \overline{\pi(u,w,v)}$ and
$h_{\ell,{\ell'}}(u,w,v) = h_{{\ell'}}(u,w,v) = h(w) =
h_\ell(\pi(u,w,v))$ are verified.  Thus $\pi: \Gamma_{\ell,{\ell'}}^*
\to \Gamma_\ell^*$ defines a projection such that
  $$\pi^*(E_\ell) = (\Gamma_{\ell,{\ell'}}, h_{\ell,{\ell'}},
  \Omega_{\ell}, \rho_\ell; L_\ell = R_\ell).$$
We have to define a partial solution $\delta: \Omega_{\ell} \to
\Gamma_{\ell,{\ell'}}^* \Omega_{{\ell'}} \Gamma^*_{\ell,{\ell'}} \cup
\Gamma^*_{\ell,{\ell'}}$ such that $\delta(L_\ell) = \beta(L_{{\ell'}})$
and $\delta(R_\ell) = \beta(R_{{\ell'}})$.  For this, we have to consider a
variable $X\in\Omega$ with $\body_\ell(\sigma(X)) \neq 1$.
 By symmetry, we may
assume that $X = x_i$ for some $1 \leq i \leq g$.  Hence $\sigma(X) =
w_0[\leftp(i),\rightp(i)]$.

Let $\alpha = \leftp(i) + |\head_\ell(\sigma(X))|$ and $\beta =
\rightp(i) - |\tail_\ell(\sigma(X))|$.  Then $\leftp(i) + \ell \leq
\alpha < \beta \leq \rightp(i)-\ell$.  Let $(u_p,w_p,v_p) \cdots
(u_q,w_q,v_q)$ be the minimal cover of $[\alpha,\beta]$ with respect
to the $\ell$-factorization.  We have $w_p\cdots w_q =
\body_\ell(\sigma(X))$.

For $\body_{{\ell'}}(X) = 1$ we have $X\in\Omega_\ell \setminus
\Omega_{{\ell'}}$ and we define $$\delta(X) = (u_p,w_p,v_p) \cdots
(u_q,w_q,v_q).$$  Then $\delta(X)\in B_\ell^*$ and $h_\ell \delta(X) =
\rho_\ell(X)$ since $\rho_\ell(X) = h(\body_\ell( \sigma(X)))$.  It is
also clear that the definition does not depend on the choice of $i$,
and we have $\delta(\overline{X}) = \overline{\delta(X)}$.

Recall the definition of $L_{{\ell'}}$.  Since $\body_{{\ell'}}(\sigma(X))
= 1$, there is a factor $f_1\cdots f_r$ of $L_{{\ell'}}$ which belongs
to $\Gamma_{{\ell'}}^*$ and $f_1\cdots f_r$ covers $[\alpha,\beta]$ with
respect to the ${\ell'}$-factorization $F_{{\ell'}}(w_0)$.  It follows
that $\delta(X)$ is a factor of $\beta(f_1\cdots f_r)$, hence
$\delta(X) \in \Gamma_{\ell,{\ell'}}^*$ by definition of
$\Gamma_{\ell,{\ell'}}$.

For $\body_{{\ell'}}(X) \neq 1$ we have $X\in\Omega_{{\ell'}}$ and we find
positions $\mu < \nu$ such that $\mu = \leftp(i) +
|\head_{{\ell'}}(\sigma(X))|$ and $\nu = \rightp(i) -
|\tail_{{\ell'}}(\sigma(X))|$.

For some $p \leq r \leq s \leq q$ we have $w_0[\alpha,\mu] = w_p
\cdots w_{r-1}$, $w_0[\nu,\beta] = w_{s+1} \cdots w_q$, and
$\body_{{\ell'}}(\sigma(X)) = w_r \cdots w_s$.  We define
\begin{displaymath}
  \delta(X) = (u_p,w_p,v_p) \cdots (u_{r-1},w_{r-1},v_{r-1}) X
  (u_{s+1},w_{s+1},v_{s+1}) \cdots (u_q,w_q,v_q).
\end{displaymath}
As above, we can verify that $\delta(X) = UXV$ with $U,V \in
\Gamma_{\ell,{\ell'}}^*$ such that $\delta(\overline{X}) =
\overline{V}\,\overline{X}\,\overline{U}$ and $\rho_\ell(X) =
h_{\ell,{\ell'}}(U) \rho_{{\ell'}}(X) h_{\ell,{\ell'}}(V)$.
Finally, $\delta(L_\ell) = L_{{\ell'}}$ and $\delta(R_\ell) =
R_{{\ell'}}$.  Hence $\delta_*(\pi^*(E_\ell)) = \beta_*(E_{{\ell'}})$.
This proves Theorem~\ref{theo3}.

\section{Conclusion}
In this paper we were dealing with the existential theory, only. For
free groups it is also known that the positive theory without
constraints is decidable, see \cite{mak84}. Thus, one can allow also
a mixture of existential and universal quantifiers, if there 
are no negations at all. Since a negation can be replaced with
the help of an extra variable and some positive rational constraint,
one might be tempted to prove that 
the
positive theory of equations with rational constraints in free groups
is  decidable. But such a program must fail: Indeed,
by \cite{marchenkov82} and \cite{durnev95} it is known that  the
positive $\forall\exists^{3}$-theory of word equations is unsolvable.
Since $\Sigma^*$ is a rational
subset of the free group
$F(\Sigma)$, this theory can be encoded in the positive theory
 of equations with rational constraints in free groups, and the 
later is undecidable, too. On the other hand, a
negation leads to a positive constraint of a very restricted type, so
the interesting question remains under which type of constraints the
positive theory becomes decidable.

\paragraph{Acknowledgments} 
The research was partly supported by the German Research Foundation
{\em Deutsche Forschungsgemeinschaft, DFG}. In addition, C.~Guti\'errez thanks 
Centro de Modelamiento Matem\'atico, FONDAP Ma\-te\-m\'a\-ti\-cas Discretas,
for financial support.

\bibliographystyle{plain}

\newcommand{\IJ}{IJ}\newcommand{\Ju}{Ju}\newcommand{\Th}{Th}\newcommand{\Yu}{Y%
u}\newcommand{\HHHH}{}\newcommand{\VVVV}{}\newcommand{\NNNN}{}

\end{document}